\documentclass[12pt, preprint]{aastex}

\usepackage{xspace}

\newcommand{\chandra}{\textit{Chandra}\xspace}
\newcommand{\xmm}{\textit{XMM-Newton}\xspace}
\newcommand{\euve}{\textit{EUVE}\xspace}
\newcommand{\asca}{\textit{ASCA}\xspace}

\begin{document}

\title{Capella Corona Revisited: A Combined View from \textit{XMM-Newton} RGS,
  \textit{Chandra} HETGS, and LETGS}

\author{
M. F. Gu,
R. Gupta,
J. R. Peterson,
M. Sako,
and
S. M. Kahn}
\affil{Department of Physics and Kavli Institute for Particle Astrophysics and
  Cosmology, Stanford University, CA 94305}

\begin{abstract}
We present a combined analysis of the X-ray emission of the Capella corona
obtained with \textit{XMM-Newton} RGS, \textit{Chandra} HETGS, and LETGS. An
improved atomic line database and a new differential emission measure (DEM)
deconvolution method are developed for this purpose. Our new atomic database
is based on the Astrophysical Plasma Emission Database and incorporates
improved calculations of ionization equilibrium and line emissivities for
L-shell ions of abundant elements using the Flexible Atomic Code. The new DEM
deconvolution method uses a Markov Chain Monte-Carlo (MCMC) technique which
differs from existing MCMC or $\chi^2$-fitting based methods. We analyze the
advantages and disadvantages of each individual instrument in determining the
DEM and elemental abundances. We conclude that results from either RGS or
HETGS data alone are not robust enough due to their failure to constrain DEM in
some temperature region or the lack of significant continuum emission in the
wavelength band of the spectrometers, and that the combination of HETGS and
RGS gives more stringent constraints on the DEM and abundance
determinations. Using the LETGS data, we show that the 
recently discovered inconsistencies between the EUV and X-ray lines of Fe
XVIII and XIX also exist in more highly charged iron ions, up to Fe XXIII, and
that enhanced interstellar absorption due to partially ionized plasma along the
Capella line of sight may explain some, but not all, of these discrepancies.
\end{abstract}

\section{Introduction}
The Capella system is one of the strongest X-ray emitting coronal sources, and
has been observed numerous times with the past and current X-ray
observatories, including both High and Low Energy Transmission Grating
Spectrometers (HETGS, LETGS) on board \chandra and the Reflection Grating
Spectrometer (RGS) of \xmm. Capella is a close spectroscopic binary with an
orbital period of 104 days and a distance of 12.9~pc \citep{hummel94}. 

The general features of the temperature distribution of the Capella coronal
plasma were reasonably well determined, even with the previous generation
X-ray and EUV observatories, equipped with limited spectral
resolutions. Observations with the \euve spacecraft have shown a continuous
differential emission measure (DEM) distribution over a temperature range of
$10^5$ to $10^7$~K \citep{dupree93}. \citet{brickhouse00} analyzed the
simultaneous observations of \euve and \asca, and concluded that the DEM is
sharply peaked near $10^{6.8}$~K. The abundances of Mg, Si, S, and Fe were
found to be consistent with solar photospheric values, and Ne was found to be
underabundant by a factor of $\sim 3$ to 4 in that analysis.

Since the launch of \chandra and \xmm, high resolution X-ray spectra have become
available, and numerous analyses of the Capella coronal X-ray emission have
appeared using all three grating instruments. The first light observations
were presented by \citet{canizares00} for HETGS, \citet{brinkman00} for LETGS,
and \citet{audard01} for RGS, which gave an overview of the spectral data
using relatively simple analysis methods. \citet{behar01} investigated in
detail the Fe L-shell line emission using the HETGS data and theoretical
calculations of the Hebrew University Lawrence Livermore Atomic Code
(HULLAC). \citet{phillips01} made detailed comparisons between the HETGS X-ray
spectra and extreme-ultraviolet emission from \euve. \citet{mewe01} presented
temperature, density and abundance diagnostics using the LETGS observation and
a line ratio based analysis method. \citet{argiroffi03} studied the structure
and variability of the X-ray corona using multiple LETGS observations and a
Markov Chain Monte-Carlo (MCMC) method for the DEM deconvolution developed by
\citet{kashyap98}. The x-ray emission are found to be constant to within a few
percent on both short and long time scales. \citet{audard03} studied the
coronal abundances and the first ionization potential (FIP) effects of several
RS CVn binaries, including Capella, using the \xmm observations.

All previously mentioned analyses have relied on a single grating
instrument in the X-ray band. \citet{desai05} combined multiple HETGS and LETGS
observations and investigated various line ratios of Fe XVIII and XIX.
Large discrepancies of a factor of two were found between the observed and
theoretical ratios involving $3\to 2$ X-ray transitions and $2\to 2$ EUV
resonance lines. It was assumed that such discrepancies reflect the
uncertainties of theoretical atomic data. However, this combined analysis
was not aimed at deriving coronal properties of Capella. 

Because of the different spectral coverage of the three grating instruments,
it is conceivable that a joint analysis of all instruments may yield more
stringent constraints on the DEM distribution and elemental abundances of the
capella corona than that of individual instruments. Although observations
with three gratings are generally not simultaneous, a joint analysis is
possible due to the lack of variability of X-ray emission as indicated by
multiple observations with individual instruments. In this paper, we present
such a combined analysis of observations with HETGS (ObsID 5040, 28 ks), LETGS (ObsID
1248, 84 ks), and RGS (ObsID 0121920101, 51 ks). Many more similar
observations exist for all instruments. The three are chosen more or less
randomly. The primary goal of the present work is to introduce our analysis
method and discuss the complementary nature of the three instruments.
The statistical quality of these individual observations are sufficient for
this purpose. We leave a systematic investigation of all available data for
future work. 

In the course of this work, we have
developed an improved atomic line database and a new DEM deconvolution
method. The new atomic line database is based on the line list of
astrophysical plasma emission code (APEC) of \citet{smith01} with the
L-shell emission lines from Ne, Mg, Al, Si, S, Ar, Ca, Fe, and Ni replaced with
calculations using the Flexible Atomic Code (FAC) developed by
\citet{gu03a}. Our new MCMC DEM deconvolution method is an adaptation of the
spatial-spectral analysis procedure of \citet{peterson06} to the analysis of
pure spectral data, and is different from the MCMC method of \citet{kashyap98}
in technical details.

In \S\ref{sec:atomic}, we describe our improvements to the atomic line
database of APEC; in \S\ref{sec:dem}, the details of the MCMC method for DEM
deconvolution are discussed; \S\ref{sec:rgs} and \S\ref{sec:hetg} present the
analysis of the \xmm RGS and \chandra HETGS observations respectively;
\S\ref{sec:rhetg} presents the joint analysis of RGS and HETGS data; In
\S\ref{sec:letg}, the additional constraints on the physical properties of the
interstellar medium along the Capella sight line imposed by LETGS data are
analyzed; we conclude with brief 
discussions of the results in \S\ref{sec:discussion}.

\section{An Improved Atomic Line Database}
\label{sec:atomic}
A complete and accurate atomic database is the cornerstone of X-ray
spectroscopy. The high resolving power of \chandra and \xmm grating
spectrometers have made it even more important because many previously
unresolved line transitions become well isolated in the new
observations. There has been a concerted effort, both experimental and
theoretical, to improve the existing atomic data in the past decade. For
example, the widely used MEKAL plasma model has gone over major revisions in
its implementation in the SPEX package. \citet{smith01} have developed a new
plasma emission code and its associated atomic data, combining the most recent
theoretical and experimental results for many important transitions,
particularly those from iron L-shell ions. This database and the code,
astrophysical plasma emission database and code (APED/APEC) are routinely used
in the analyses of \chandra and \xmm grating data. 

Despite the numerous improvements and greatly enriched line list in APED/APEC
over its predecessors, there are some deficiencies in its most recent
version. Firstly, the collisional ionization balance used in the code are
based on the recombination and ionization rates compiled by
\citet{mazzotta98}. \citet{gu03b} showed that there are some systematic
problems with dielectronic recombination (DR) rate coefficients used there for
some L-shell ions of astrophysically abundant elements, including iron. The
ionization fractions calculated using the new DR rate coefficients of
\citet{gu03b} can be different from that of \citet{mazzotta98} by as much as a
factor of two for some ions. Secondly, the line emissivities of L-shell ions
in APED/APEC are mostly based on the distorted-wave calculations of HULLAC,
and do not take into account the roles of resonances. \citet{gu03a} and
\citet{doron02} demonstrated that not only do resonances play an important role in
forming X-ray lines for L-shell iron ions, but also the DR of next higher charge
state contributes to the line emissivities. Thirdly, the transition
wavelengths of L-shell ions have primarily relied on HULLAC calculations
except for a subset of lines from the iron ions where experimental
measurements of \citet{brown98} and \citet{brown02} are used. HULLAC
wavelengths are known 
to have limited accuracy insufficient for the analysis of \chandra and \xmm
grating spectra. Lastly, the L-shell line models for less abundant elements
such as Si, S, Ar, and Ca are not as sophisticated as that for Fe, and do not
contain as many transitions.

We have modified the APED/APEC line list by replacing all L-shell emission
lines of Ne, Mg, Al, Si, S, Ar, Ca, Fe, and Ni with our own calculations using
the atomic data produced by the Flexible Atomic Code. For the $n=3\to n=2$
lines of Fe ions, we use the data of \citet{gu03a}, which includes resonance
excitation, recombination and ionization processes as population
mechanisms. For other lines of Fe and other elements, we have calculated
direct excitation only models including levels up to $n=12$. The ionization
fractions of all L-shell ions are calculated with the DR rate coefficients of
\citet{gu03b} in our new models. We have used the many-body perturbation
calculation of \citet{gu05} for the wavelengths of $3\to 2$ lines of Fe and Ni
ions, which have been checked against the measurements of
\citet{brown98} and \citet{brown02} for strong Fe lines and deemed accurate to
within a few 
m{\AA}. For the L-shell lines of S, and Ar, we have replaced the
theoretical wavelengths with experimental ones for those identified in the
measurements of \citet{lepson03} and \citet{lepson05}. We calculated our
emissivities in 
the low density limit, which is appropriate for the majority of lines at stellar
coronal densities in general, and for Capella in particular. Some lines in the
lower Z ions may have been affected by density effects. We estimate that such
effects are not severe at the relatively low density of the Capella corona, and do
not attempt to correct for them.

\begin{figure}
\epsscale{1.0}
\plotone{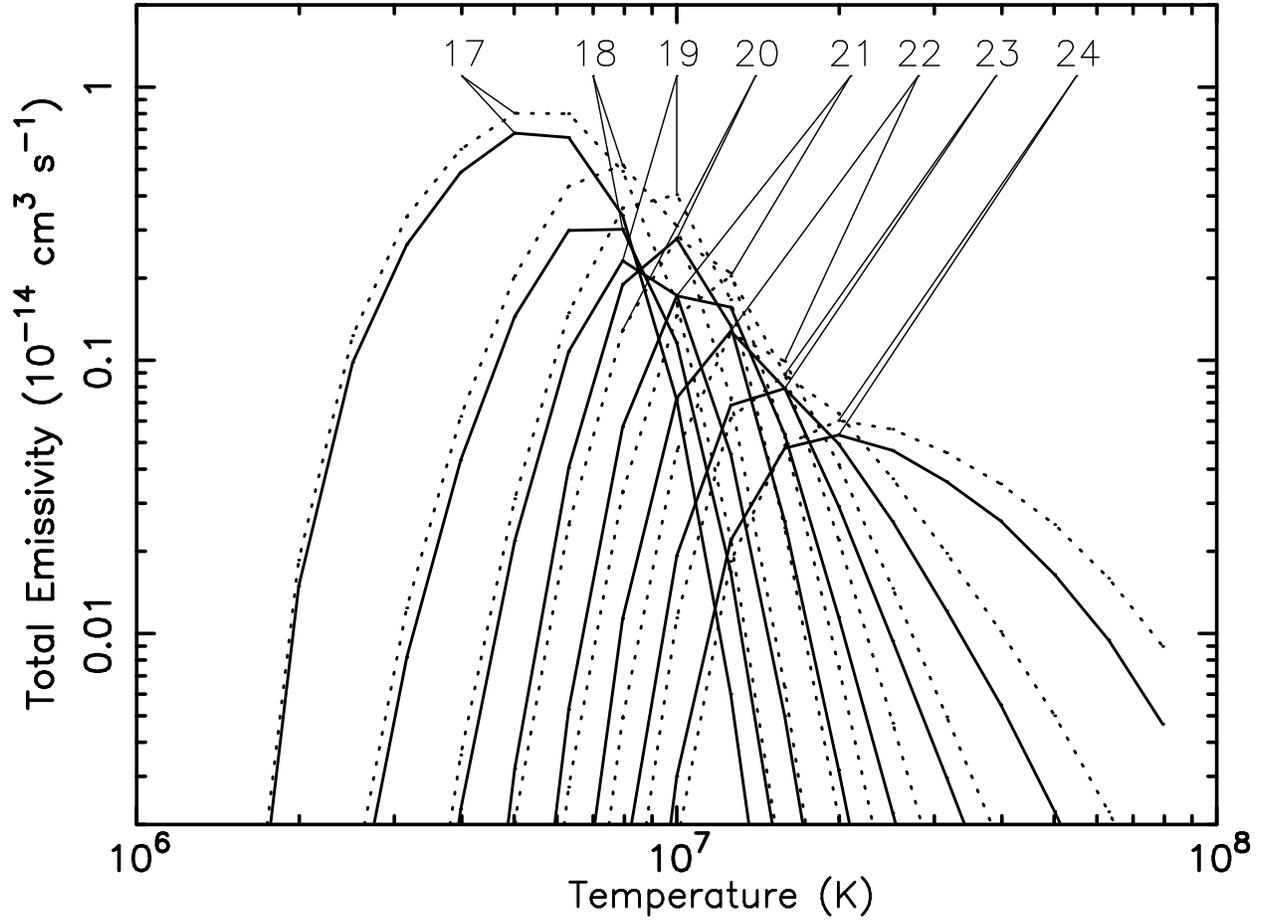}
\caption{\label{fig:emiss}Total emissivities of Fe L-shell ions in the
  5--20~{\AA} region. The solid
  lines are from the original APED/APEC database; the dotted lines are from
  our new line list. The numbers label Fe charge states, 17 for Fe XVII, 18
  for Fe XVIII, and so on.}
\end{figure}

\begin{figure}
\epsscale{1.0}
\plotone{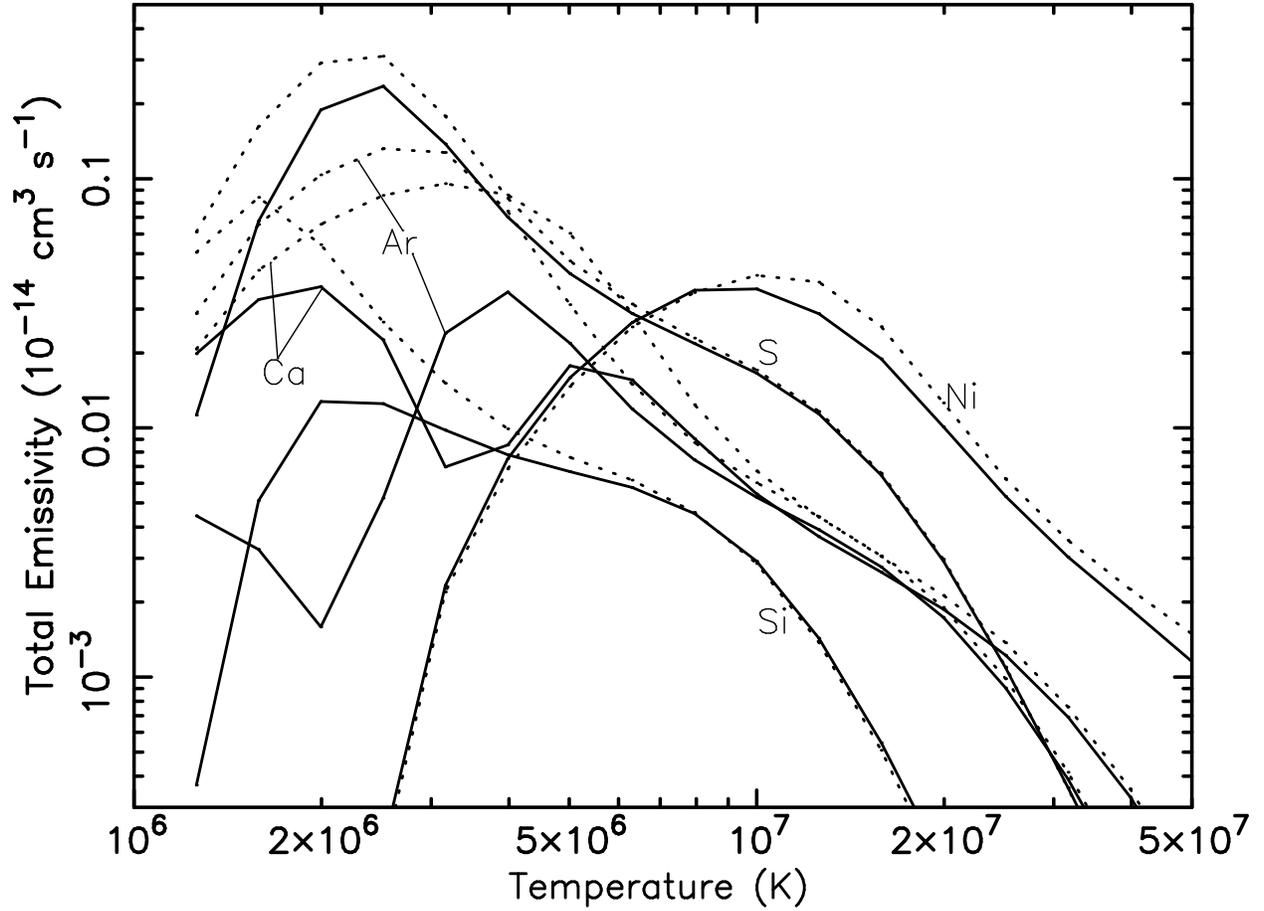}
\caption{\label{fig:emiss1} Total emissivities of L-shell ions of Si, S, Ar,
  Ca, and Ni in the 10--40~{\AA} region, summed over all charge states. The solid
  lines are from the original APED/APEC database; the dotted lines are from
  our new line list.}
\end{figure}

The result of our modification to the APED/APEC line list is a vastly enlarged
line model with more accurate wavelengths and emissivities for L-shell
ions. The new line list 
is stored in the same FITS format as the original one, and can be used in
any analysis software that already incorporates APED/APEC plasma
models. To illustrate the main differences between our new line list and the
original APED/APEC list, we show the comparison of total emissivities of Fe
L-shell ions in the 5--20~{\AA} spectral region as 
functions of temperature in Figure~\ref{fig:emiss}. The emissivities in our
new database are generally larger than those of APED/APEC, reflecting the
inclusion of resonances and recombination contributions to the level
population and the modification to ionization equilibrium
fractions. Enhancements of more than a factor of two are seen at certain
temperature ranges for some ions. For Fe XVIII and XIX, even the differences
in peak emissivities reach a factor of 1.7. In Figure~\ref{fig:emiss1}, we
show the total emissivities of L-shell ions of Si, S, Ar, Ca, and Ni in the
5--40~{\AA} spectral region as
functions of temperature. It is seen that our new emissivities are also
generally larger than the original APED/APEC database. The differences in the
Ni emissivities are relatively small, while for other elements, they can be
quite larger at temperatures of 1--5$\times 10^6$~K. We also note that our
new data does not exhibit the irregular temperature variation for Ar and Ca
emissivities, as does the APED/APEC. In the next section, we will
examine the differences between the two 
line models in further detail when they are applied to analyze some simulated data
using our DEM deconvolution method.

\section{A New Differential Emission Measure Deconvolution Method}
\label{sec:dem}
Differential emission measure (DEM) distribution is one of the fundamental
properties of stellar coronae. It characterizes the temperature structure
of the coronal plasma, and is the main objective of X-ray spectroscopic
analyses. Many techniques have been developed to derive DEM from X-ray or EUV
spectroscopic data. For low resolution instruments such as the solid state
detectors of \asca, simple multi-temperature model fitting methods are usually
favored since individual emission lines cannot be resolved. For high
resolution instruments, the first step in deriving DEM in almost all existing
methods is to identify individual
lines from different ions and measure their fluxes. In the simplest method,
one then assumes that each ion emits at the temperature of peak
emissivity. The amount of emission measure at that temperature is then derived
by dividing the flux by the peak emissivity, assuming known abundance of the
emitting element. The discrepancy between the emission measure derived with
different elements but at similar temperatures can be used to infer the
deviation of the elemental 
abundances from the assumed values. In more advanced techniques, one may try
to infer the DEM and elemental abundances simultaneously by matching the model
and measured line fluxes and continuum level. The actual matching procedure
may take the form as $\chi^2$ fitting or MCMC simulations.

Using the measured line flux as the starting point has its advantages and
disadvantages. The main advantage is that accurate theoretical wavelengths of
transitions and wavelength calibration of the instruments are not critical as
long as lines can be reliably identified, 
while the disadvantage is that it requires the lines to be well resolved to be
useful in this procedure. However, even with the high resolution of \chandra
and \xmm grating spectrometers, many lines still blend with each other and
their fluxes are difficult to measure individually. With the wavelengths of
all major strong lines known to a high accuracy either through laboratory
measurements or many-body perturbation calculations, we find that a global
fitting method to the raw spectrum is feasible and is our preferred method
because it takes into account all lines and continuum emission on equal
footing. The possible effects of wavelength uncertainties in the database will
be discussed in more detail later in this section. In our analysis, we also
work with the raw spectra without rebinning. The problem of low photon counts
are taken care of by using the Poisson likelyhood throughout the entire
spectral region.

One still needs to decide how the DEM is to be parameterized and the actual
fitting technique to be used. Clearly, the parameterization of the DEM and the
abundances of major cosmic elements requires a fairly large parameter
space. The traditional minimization methods are not only inefficient but
often easily trapped in local minima. The MCMC method is a natural choice for
such problems. In addition to using a global fitting method instead of line flux
based techniques, we also choose a different parameterization framework from the
MCMC method of \citet{kashyap98}. In the method of \citet{kashyap98}, the DEM
distribution is discretized on a suitable temperature grid, and the emission
measure value at each temperature bin is treated as a parameter. Our method is
an adaptation of the more sophisticated technique of smoothed particles
inference for spatial-spectral analysis \citep{peterson06} to the pure
spectral data. In this method, the plasma is represented as a set of smoothed
particles with identical emission measures at different temperatures. In the
spatial-spectral analysis, each particle has a Gaussian shape in spatial
extension. In the pure spectral case as applicable to the stellar coronae, we
ignore the spatial extension, and each particle becomes a
basic luminosity block at certain temperature. In other words, the
DEM is represented by a multi-temperature model
\begin{equation}
\frac{dEM}{d\log T} = \sum_i\eta\delta(\log T-\log T_i),
\end{equation}
where the emission measure normalization at each temperature, $T_i$, is the
same, and given by the global parameter $\eta$. Unlike the parameterization of
discretizing the DEM over a fixed temperature grid,
here the temperatures of luminosity particles are taken as free
parameters. The peaks and valleys in the DEM therefore correspond to greater
or lesser concentrations of these particles in certain temperature regions. 
Initially, the temperatures
of the particles are drawn randomly from a uniform distribution on the
logarithmic scale over a suitable
temperature range. They are then iterated in the MCMC process along with all
other required parameters for the plasma, such as the interstellar absorption and
elemental abundances. When the Markov Chain has reached equilibrium, the
histogram of the set of temperatures in each iteration provides an estimate of
the DEM distribution. The average of this histogram along the equilibrium chain gives
the smoothed DEM with greatly reduced statistical noise, and is taken as the final
result. The variance of the histogram is taken as the statistical uncertainty of
the emission measure in each temperature bin. Other plasma parameters and
their uncertainties are
similarly estimated by taking averages and variances along the Markov
chain. We implemented our method as a custom-built fit method in the
Interactive Spectral Interpretation System (ISIS, \citet{houck00}), and the
analysis presented below are all carried out using ISIS.

The main advantage of this parameterization procedure is that the DEM
distribution is naturally and adaptively smoothed over a temperature scale
comparable to the 
width of emissivity profiles of individual lines. Such smoothing is realized
by the random walk of the temperature particles and the averaging of
temperature histograms over a long Markov chain in equilibrium. If one parameterizes
the DEM over a fixed temperature grid, some regularization technique must be
employed to ensure the smoothness of the distribution, and such smoothing must
be predetermined regardless of how much is actually required by the data and
model. Even though our new parameterization requires a large number of
parameters as we typically use 100-200 temperature particles to represent the
DEM distribution, it is found to perform quite robustly in our tests. In
particular, it always finds the DEM solution consistent with the data and
without overly sharp features, as we demonstrate below.

Another advantage of our parametrization is the relatively small correlations
between all parameters, except for the overall normalization parameter, $\eta$, and
the overall absolute abundance, which are the only two that are highly
correlated. In order to avoid dealing with this high correlation, we stop
iterating the $\eta$ parameter after a certain number of iterations when the MCMC
chain has entered the equilibrium state, and a reasonable estimate of $\eta$
has been found. This way, we only adequately sample
the parameter space that defines the shape of the DEM and the relative
abundances. Therefore the statistical uncertainties of abundances given by our
analysis are only meaningful in a relative sense.

\begin{figure}
\epsscale{1.0}
\plotone{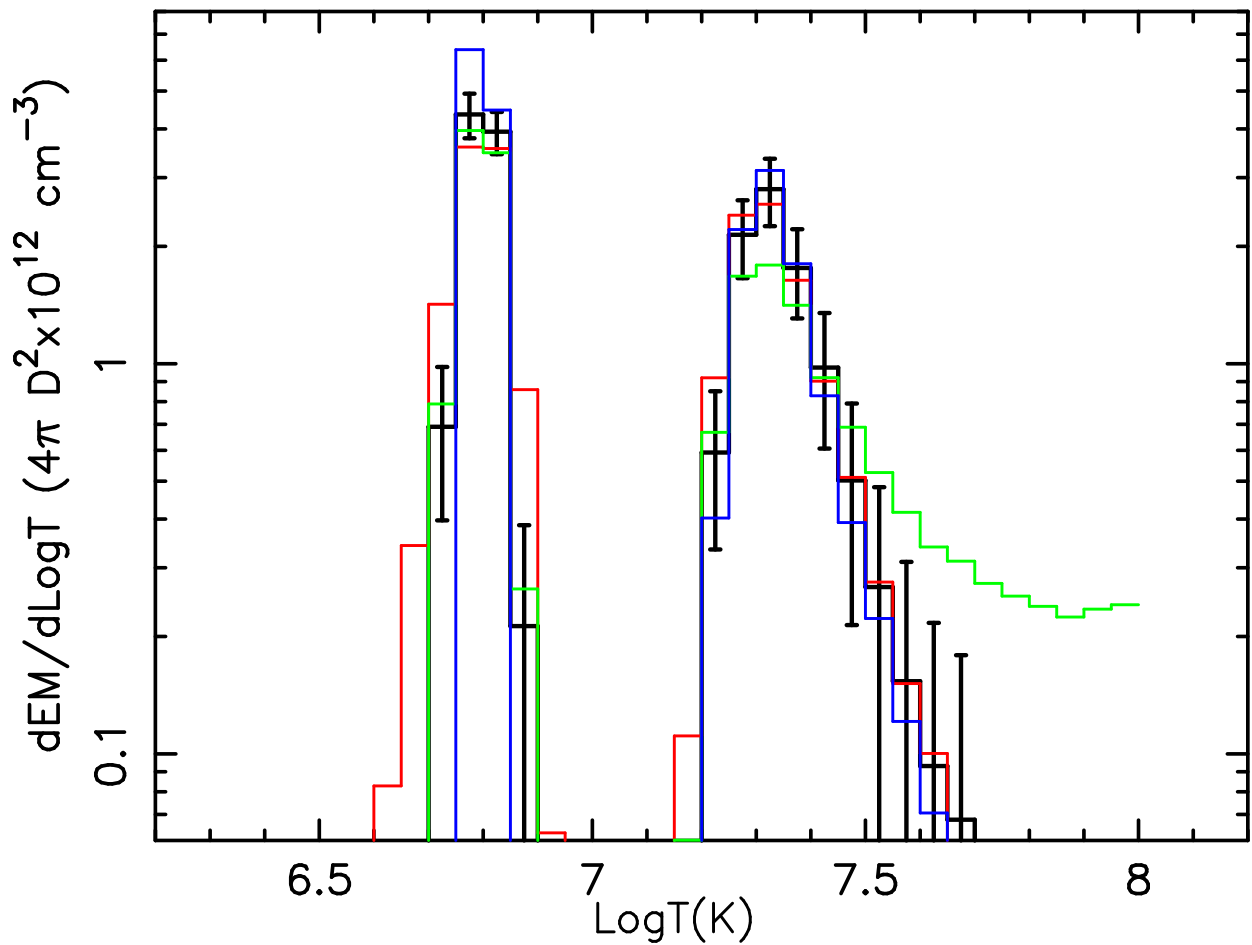}
\caption{\label{fig:sim1dem}The reconstructed DEM for the first test case with
two temperature components at $\log(T) = 6.8$ and $7.3$; the black histogram
with error bars is the reconstruction from the joint fit of MEG and RGS data;
the green histogram is the reconstruction from the RGS data alone; the red
histogram is the reconstruction from the MEG data alone; and the blue
histogram is the reconstruction using the original APED/APEC database. The
error bars represent $1\sigma$ statistical uncertainties.}
\end{figure}

In our first test, we simulate a two-temperature thermal plasma with $\log(T)
= 6.8$ and $7.3$. The total emission measure in the two components are equal
and is $2\pi D^2\times 10^{12}$~cm$^{-3}$, where $D$ is the distance of the
source in cm. We assume solar abundances and no interstellar absorption in the
model. The spectrum is then folded through instrument responses to
generate fake data for 60 ks observations for both HETGS and RGS, which are
analyzed by the present MCMC method. The total emisssion measure and
exposure time chosen here make the spectra in both HETGS and RGS comparable
to the Capella data analyzed in this work. During the 
analysis, we use 100 temperature particles. The abundances of all
heavy elements are tied as a single parameter but are allowed to
vary. Including the overall normalization factor, $\eta$, a total of 102
parameters are therefore required to characterize the plasma model. 
The MCMC chain is 2000 iterations long, and 
we exclude the first 500 iterations as the burn-in period. Through visual
inspection of key parameters, we determine that 500 burn-in iterations are
more than enough to allow the MCMC chain to enter the equilibrium state. 
The test is
designed to illustrate how the method behaves when the underlying DEM has two
sharp peaks. The MCMC analysis is performed three times, on MEG data alone, on
RGS data alone, and jointly on MEG and RGS data. The three versions of the
reconstructed DEM are shown in Figure~\ref{fig:sim1dem} with red, green and
black histograms. It can be
seen that the two temperature peaks are resolved in our reconstructions, but
each has a finite width. This is because the line emissivity profiles that
constrain the DEM at the relevant temperatures have finite widths and it is
simply not possible for any method to resolve structures sharper than these
emissivity profiles allow. We also note that the high temperature peak
is broader than the low temperature one. This is because the lines that
constrain the high temperature DEM are mainly from K-shell ions, and have
broader emissivity profiles. The high temperature peak reconstructed from RGS
data is broader than those from MEG or the combination of MEG and RGS,
reflecting the short wavelength cutoff of the RGS effective area. The derived
abundance is $1.28\pm 0.05$ from MEG, $1.21\pm 0.03$ from RGS, and $1.13\pm
0.02$ from the combination of MEG and RGS. The uncertainties are quoted at
90\% confidence level, and are calculated from the variance of the MCMC
chain. These abundance values range from 10--30\% larger  
than the solar value used in the generation of the simulated data.
However, this overestimation of abundance is offset by the corresponding
underestimation of total emission measure in the reconstructed DEMs, and
therefore the product of emission measure and the abundance
agrees with the input model very well. This mismatch is due to the fact that for this
test case, there is only a weak continuum in both MEG and RGS wavelength
bands. It is the ratio of line to continuum emission that constrains the
abundance value. With weak continuum, it is difficult to reproduce the total
emission measure and abundance independently, even though the shape of the DEM
and the product of the total emission measure and the abundance can be reliably
determined. The uncertainties associated with this difficulty are not
represented in our analysis, because we stop iterating the overall normalization
parameter, $\eta$, when a reasonable estimate has been found.
To verify that the deviations in the derived absolute abundance values are in
fact consistent with the statistical properties of the data, we evaluate the
maximum likelyhood statistics as an indication of the goodness of fit. This
statistics is defined as
\begin{equation}
X^2 = -2\ln(L),
\end{equation}
where $L$ is the Poisson likelyhood. 
This definition ensures that $X^2$ approches the conventional $\chi^2$ value
when all bins have sufficient counts to be approximated by normal
distributions. Take the reconstruction from MEG as an example, the $X^2$ value
for the final model is 23651 in the 3 to 27~{\AA} range of both orders of
spectra, with 9600 total number of bins. If we fix the abundance at the solar
value and repeat the MCMC analysis, the final $X^2$ value is 23618, i.e., 33
smaller than the case with a varying abundance. However, the variance of $X^2$
for random data are expected to be $2\nu$, where $\nu = 9600$. In fact, by
generating many copies of the random data, we 
verify that $X^2$ is well approximated by a normal distribution with 
standard deviation of $\sqrt{2\nu} \sim 138$, and mean of 23297. In other words, our
reconstruction with overestimated abundance and underestimated total emission
measure does not differ statistically from the model with the correct
abundance, although both deviate from the ``perfect'' model by about 2.5$\sigma$.

To illustrate the effects of differences between our new atomic line database
and the original APED/APEC list, we carry out a further analysis. We take the
joint MEG and RGS data generated using the new atomic line list, and  analyze
them with the original APED/APEC database. The reconstructed DEM is also shown in
Figure~\ref{fig:sim1dem} as the blue histogram, and the corresponding
abundance is 1.15. Clearly, for this test case, the results are not very
different from the one obtained with the ``correct'' atomic line list, except
that the low temperature peak appears to be much narrower now. However, the
$X^2$ value of the fit is larger than that of the ``correct'' model by 3070,
which is much larger than the expected statistical deviation.

\begin{figure}
\epsscale{1.0}
\plotone{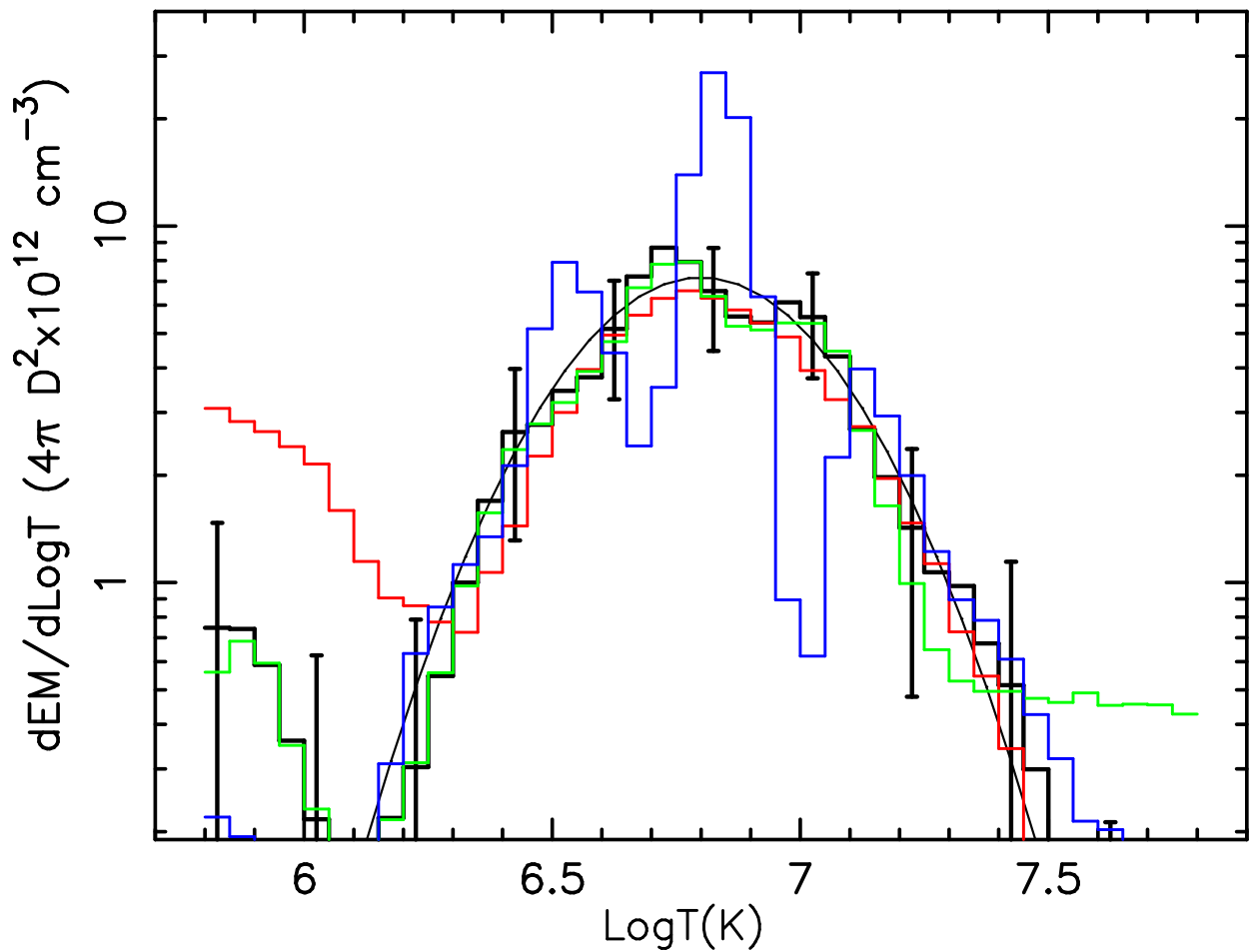}
\caption{\label{fig:sim2dem}The reconstructed DEM for the second test
  case. The black smooth line is the input model; the black histogram with
  error bars is the
  reconstruction from the joint fit of MEG and RGS data; the green histogram
  is the reconstruction from the RGS data alone; the red histogram is the
  reconstruction from the MEG data alone; and the blue histogram is the
  reconstruction using the original APED/APEC database. The error bars
  represent $1\sigma$ statistical uncertainties, and are plotted every fifth bin
  to make the figure less cluttered.}
\end{figure}

\begin{figure}
\epsscale{1.0}
\plotone{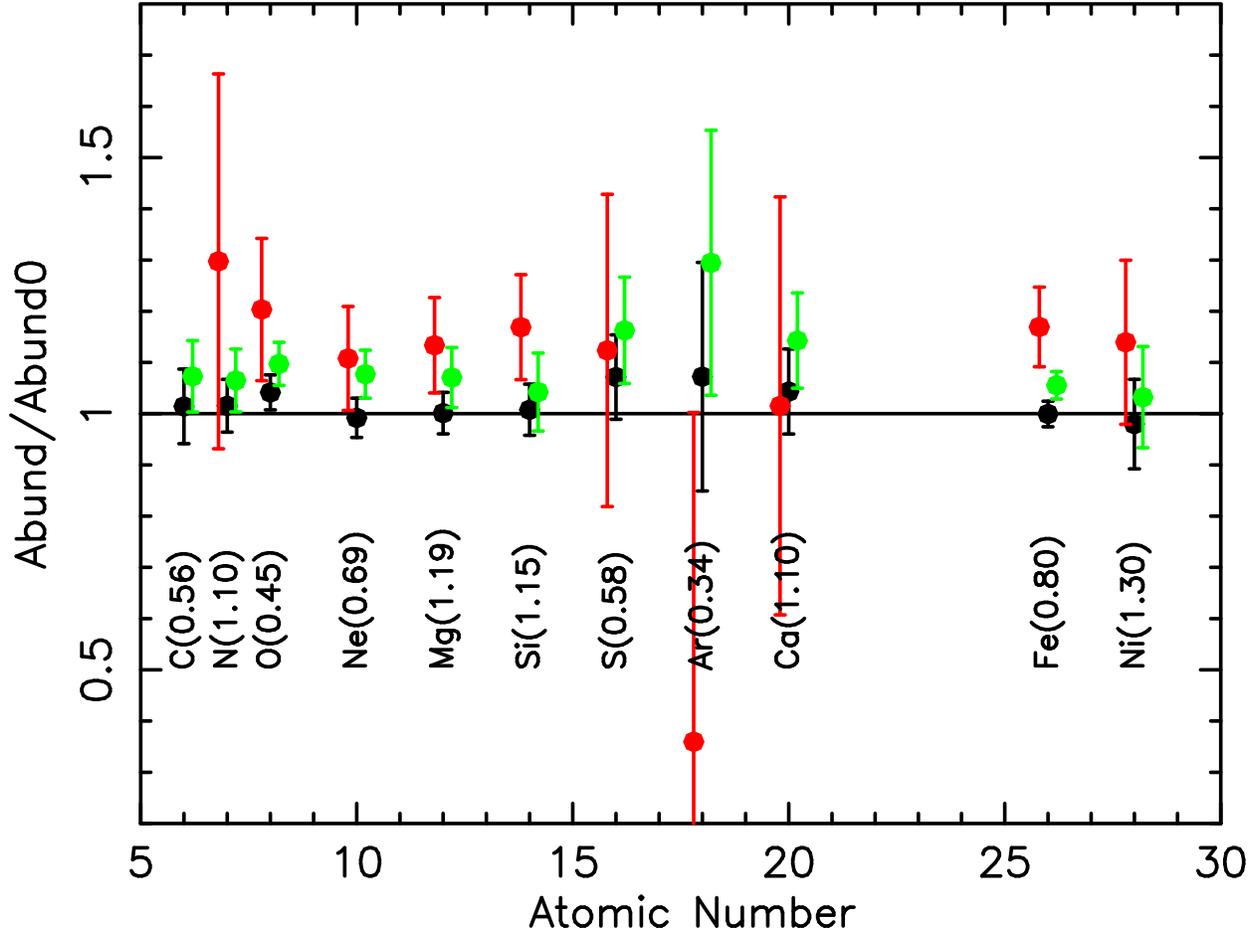}
\caption{\label{fig:sim2ab}The derived abundances relative to the input model
  in the second test case. The black symbols are the results from the joint
  fit of MEG and RGS data; the green symbols are the results from the RGS data
  alone; and the red symbols are the results from the MEG data alone. The
  error bars represent statistical uncertainties at 90\% confidence level. The
numbers in the parentheses are the abundances of the input model in solar
photospheric units \citep{anders89}.}
\end{figure}

In our second test, we simulate a thermal plasma with a broad DEM
distribution. The distribution has a Gaussian profile in $\log(T)$, which is
centered at 6.8 and has a standard deviation of 0.25. This DEM somewhat
resembles the main peak in the DEM reconstructed from the Capella data in
\S\ref{sec:rhetg}. The total emission 
measure is $1.8\pi D^2\times 10^{13}$~cm$^{-3}$, which is also comparable to the
Capella corona. Fake MEG and RGS spectra are generated with exposure times
identical to the actual Capella observations used in the present
work. Therefore the statistical quality of the simulated spectra are similar to the
Capella data analyzed in this work. We also adopt
the elemental abundances derived from the Capella data, and
allow all abundances to be free parameters in the reconstruction. Like in the
first test case, the
MCMC analysis is performed three times, on MEG data alone, on RGS data alone,
and jointly on MEG and RGS data. The three versions of the reconstructed DEM are
shown in Figure~\ref{fig:sim2dem}, and the derived abundances are shown in
Figure~\ref{fig:sim2ab}. It is seen that the three reconstructions are
generally consistent with each other, and with the input Gaussian
distribution. MEG data does not constrain the DEM below $\log(T)=6.3$, and
the RGS data produces excessive emission measure above $\log(T)=7.4$. As in the first
test case, due to the lack of significant continuum emission, both MEG and RGS
data give a total emission measure slightly smaller than the 
input model, while the abundances are slightly above the input values. 
MEG data also gives larger uncertainties to the derived abundances, with
several elements practically unconstrained.
The joint fit of MEG and RGS data gives best constrained abundance values,
which are consistent with the input model. 

The effects of using the original APED/APEC line list to analyze the spectra
simulated with our new atomic data are also investigated for this test
case. The blue histogram in Figure~\ref{fig:sim2dem} shows the resulting DEM
reconstruction from the combination of MEG and RGS data. In this case, we
notice that the reconstructed DEM gives sharper structures and suggests
multiple peaks. The deviation from the ``correct'' model is much larger in
this case than in the first test. We believe that it is due to the
irregular temperature variations of emissivities for some Fe L-shell
ions. When the original APED/APEC line list is used to analyze real
astrophysical spectra, we also notice that it always 
generates more peaks than if we use our new atomic line database. Since the
fits using our new line list always provide significantly higher qualities, we
conclude that our new database is more realistic than the old one.

\begin{figure}
\epsscale{1.0}
\plotone{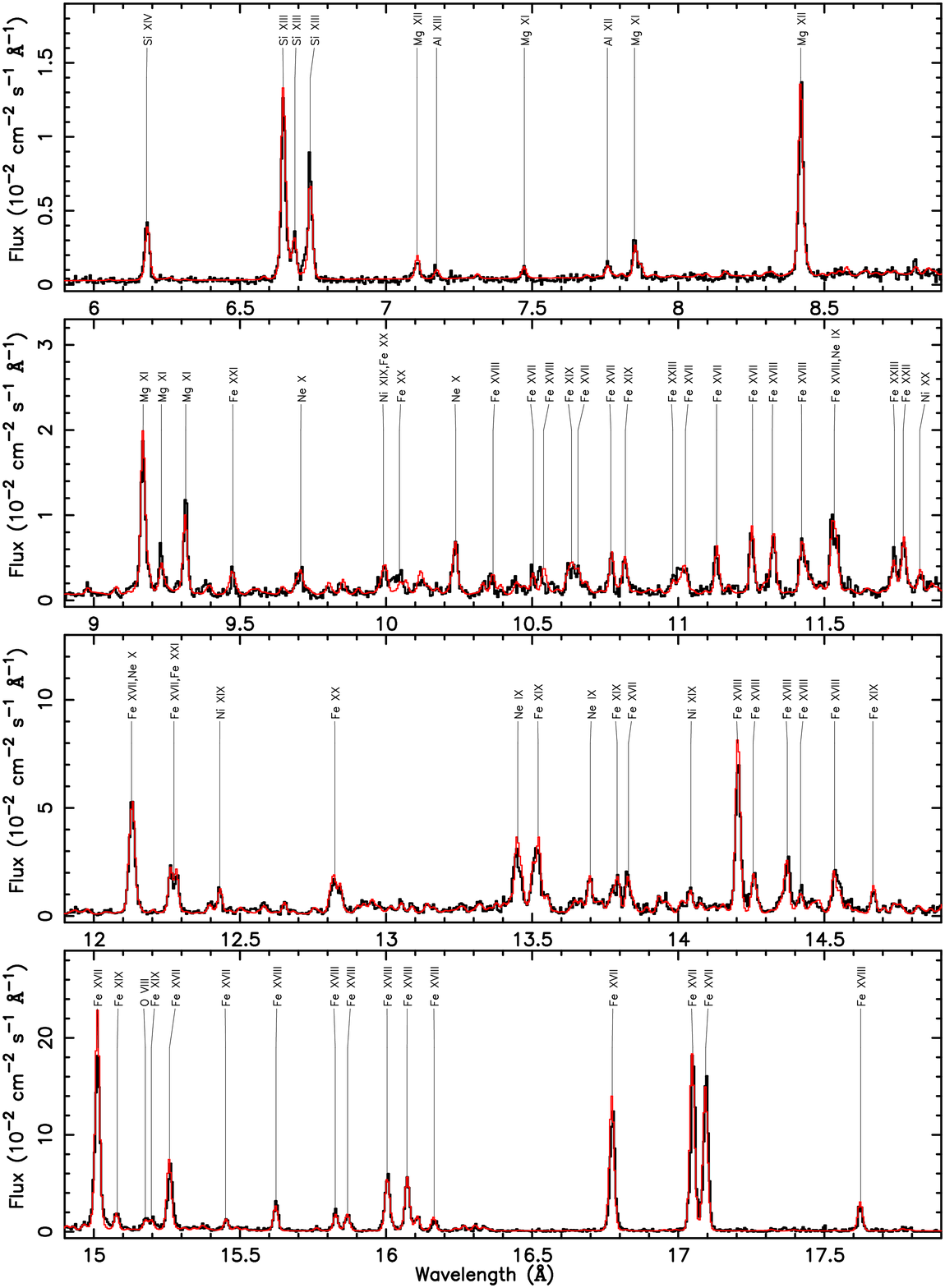}
\caption{\label{fig:hetgsp}The comparison of measured and modeled MEG spectrum
of the Capella corona in the 6--18~{\AA} region. The black line is the sum of
$\pm 1$ orders of the MEG spectra; the red line is the model using the DEM and
abundances reconstructed from the HETGS data alone.}
\end{figure}

Another major concern in a global fit method such as the one used here is
whether uncertainties in wavelengths may bias the results. In our new
database, the wavelengths of $3\to 2$ lines of L-shell ions of Fe and Ni are
obtained with a many-body perturbation calculation \citep{gu05}, whose
accuracy has been verified by comparison to the laboratory measurements of
\citet{brown98} and \citet{brown02}. We are confident that wavelengths errors
for these lines are on the order of 5~m{\AA} or less. For the strongest
transitions of S and Ar L-shell ions, the wavelength measurements of
\citet{lepson03} and \citet{lepson05} are used. The wavelengths of high-$n$
transitions of Fe ions may have slightly larger uncertainties in our
database. However, these lines are generally weaker and provide less
constraints in the DEM reconstruction. The overall good quality of fit
shown in Figure~\ref{fig:hetgsp} corroborates our error estimation. To
investigate the effects of such small wavelength uncertainties on the
results, we constructed an atomic line list by perturbing the wavelengths with
Gaussian random errors of 5~m{\AA} standard deviation, and used that to analyze
the simulated data. In both test cases discussed above, the resulting
reconstructions are the same as the ones obtained with unperturbed
wavelengths to within a few percent for both DEMs and abundances.

These two tests illustrate that when the underlying DEM has sharp features, our
method generates a smoothed reconstruction consistent with the line emissivity
profiles, and when the DEM is a continuous distribution, our reconstruction
agrees with the true DEM very well when the MEG and RGS data are
used jointly. However, such good reconstruction quality depends on the assumption
that one has the perfect knowledge of atomic database and instrument
calibration. In the analysis of real astrophysical spectra, the unknown
uncertainties in these factors may lead to systematic errors in the
reconstruction. 

\section{\xmm RGS Data}
\label{sec:rgs}
We first apply the above method to the \xmm RGS data. As in the test cases, a
total of 100 
temperature particles are used in the MCMC analysis. The temperatures are
restricted in the $10^{5.8}$ and $10^{7.8}$~K range. Initially, the temperatures
are distributed uniformly on the logarithmic scale. The abundances of C, N, O,
Ne, Mg, Si, S, Ar, Ca, Fe, and Ni are allowed to vary in the fitting, and
the interstellar absorption is fixed at the value determined by
\citet{linsky93}, which is $N_H=1.8\times 10^{18}$~cm$^{-2}$. The elemental
abundances in our analysis are measured in units of solar photospheric values
\citep{anders89}. RGS1 and RGS2
spectra are jointly fit. Including the overall normalization factor, a total
of 112 parameters are needed to characterize the plasma model.

\begin{figure}
\epsscale{1.0}
\plotone{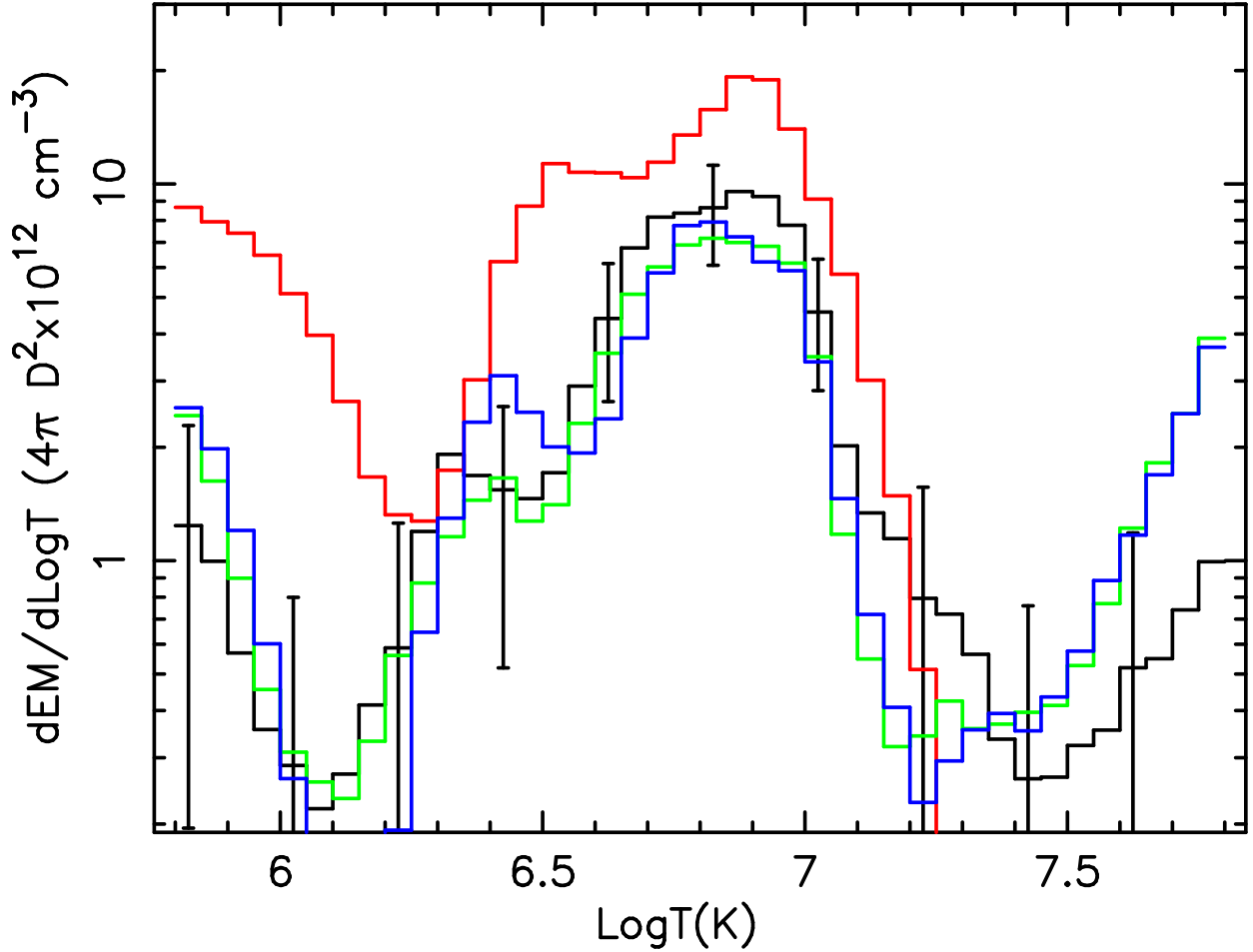}
\caption{\label{fig:dem}The reconstructed DEM for the Capella corona. The
  black histogram with error bars is the reconstruction from the joint fit of
  HETGS and RGS data; the green histogram is the reconstruction from the RGS data
  alone; the blue histogram is the reconstruction from the RGS data with the
  L-shell emission of intermediate-Z elements (Si, S, Ar, and Ca) excluded; and
  the red histogram is the reconstruction from the HETGS data alone. The error bars
  represent $1\sigma$ statistical uncertainties, and are plotted every fifth bin
  to make the figure less cluttered.}
\end{figure}

\begin{figure}
\epsscale{1.0}
\plotone{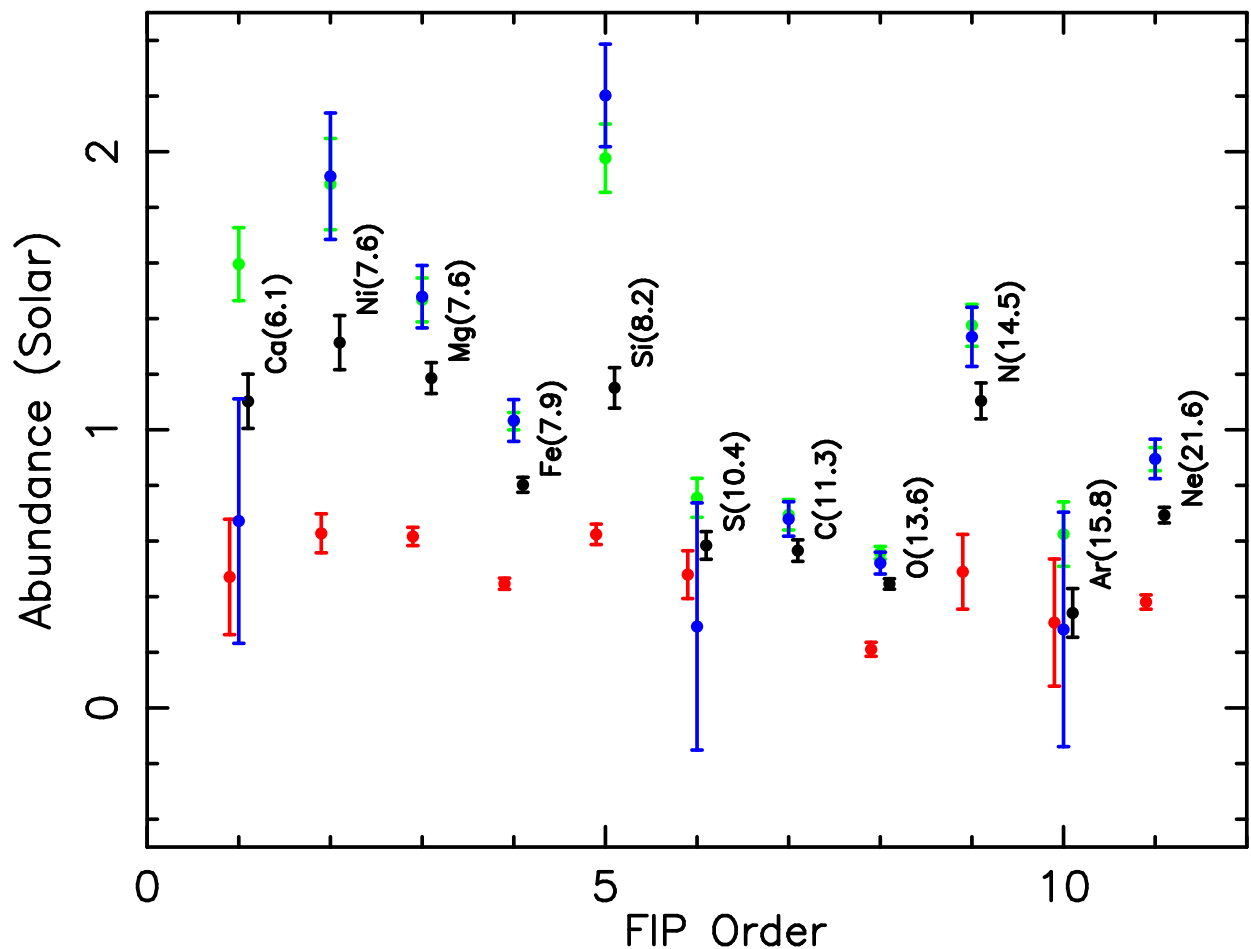}
\caption{\label{fig:ab}The derived abundances in solar photospheric units of
  \citet{anders89} for the Capella corona. The black symbols are results of
  the joint fit of HETGS 
  and RGS data; the green symbols are results of the RGS data; the blue
  symbols are results of the RGS data with the L-shell emission of
  intermediate-Z elements (Si, S, Ar, and Ca) excluded; and the red symbols
  are results of HETGS data alone. The numbers in the parentheses are the
  first ionization potentials of the elements in eV. The error bars represent
  statistical errors at 90\% confidence level.}
\end{figure}

\begin{figure}
\epsscale{1.0}
\plotone{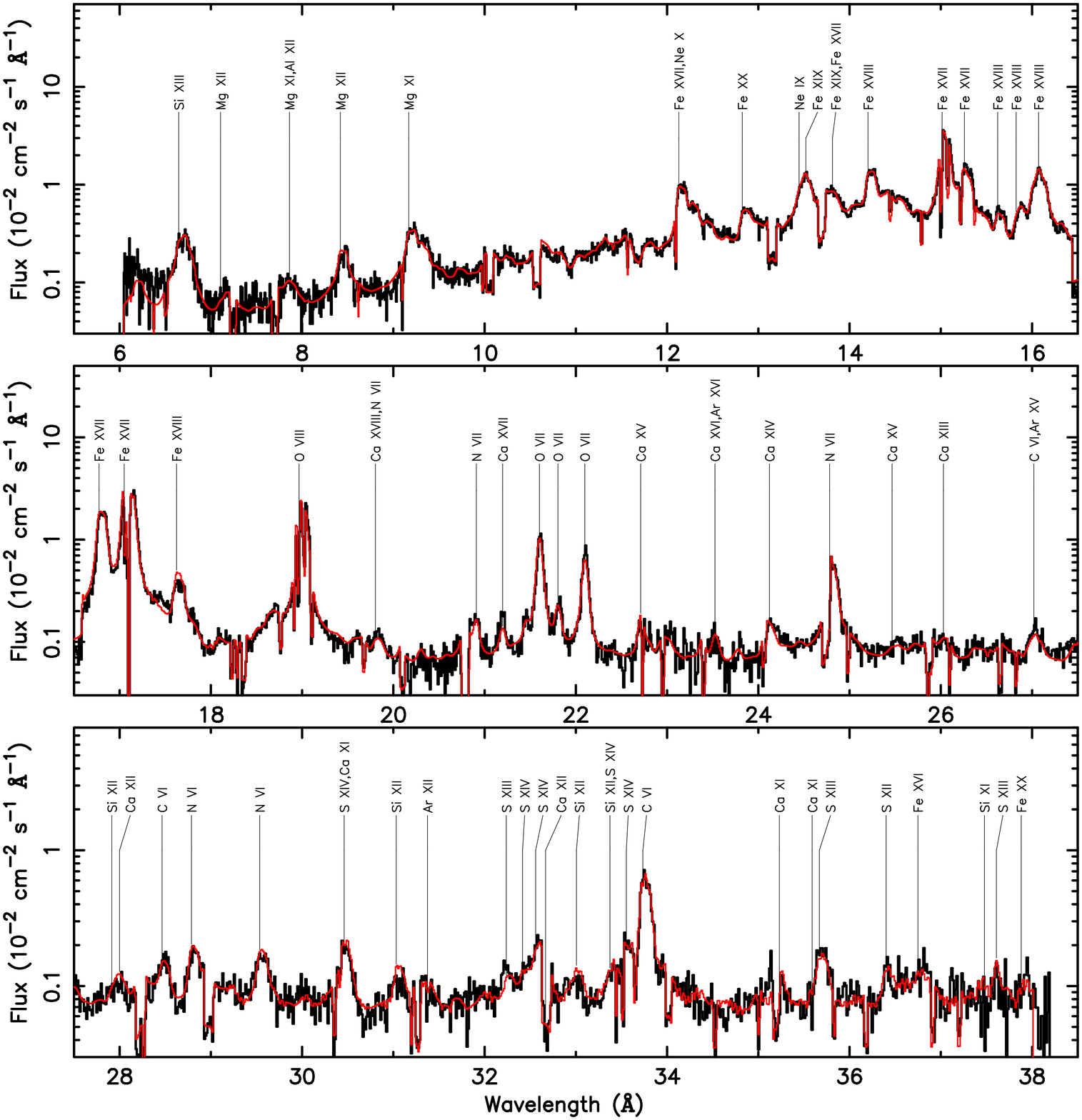}
\caption{\label{fig:rgssp}The comparison of measured and modeled RGS spectrum
  of the Capella corona. The black line is the sum of RGS1 and RGS2 data; the
  red line is the model using the DEM and abundances reconstructed from the
  RGS data alone.}
\end{figure}

Because Capella is a bright source and \xmm RGS has decent effective area at
long wavelengths, numerous L-shell lines of intermediate-Z elements, such as
Si, S, Ar, and Ca are clearly detected. The L-shell ions of these elements are
present at relatively low temperatures, while their K-shell counterparts are
generated in higher temperature regions. Assuming that the same elements in
different regions have the same abundances, the presence of both L-shell and
K-shell lines in the spectrum provides important constraints on the emission
measure of low and high temperature regions. Without these L-shell lines, the
lower temperature DEM is primarily constrained by K-shell lines of low-Z
elements, while higher temperature DEM is constrained by L-shell lines of
Fe. It would therefore be more difficult to determine the ratio of low and 
high temperature DEMs without knowing the abundance ratios of low-Z elements
and Fe. Moreover, for some elements, such as S, Ar, and Ca, the K-shell
lines are not detected by RGS, and their abundances are only constrained by
the L-shell lines. To illustrate the effects of these L-shell lines, we
carried out two fits. One includes the entire spectral region from 6 to
38~{\AA}, and the other excludes some regions where strong L-shell lines of
Si, S, Ar, and Ca are identified, primarily above $\sim 20$~{\AA}. The
resulting DEM reconstructions are shown as the
green and blue lines in Figure~\ref{fig:dem}, respectively. The abundance
measurements are shown as green and blue symbols in Figure~\ref{fig:ab}. The
summed RGS1 and RGS2 spectrum 
and the best-fit model are shown in Figure~\ref{fig:rgssp}. The reconstructed
DEMs are 
dominated by a broad peak near $\log(T)=6.8$ and a smaller secondary peak near
$\log(T)=6.4$. It is seen that the main
differences between the two DEMs are the secondary peak in the temperature
region of $\log(T)$ between 6.3 and 6.6, which is where most L-shell ions of
intermediate-Z 
elements are formed. The abundance measurements in the two fits are generally
consistent with each other, except for S, Ar, and Ca. The uncertainties of
these abundances in the fit with intermediate-Z L-shell regions excluded are
very large, and their values are practically unconstrained. The Si abundance
is still well determined because its K-shell lines fall within the RGS wavelength
band. However, the RGS effective area calibration at the Si K-shell region
is quite uncertain, making the derived Si abundances unreliable. In
Figure~\ref{fig:ab}, the abundances are plotted against the first
ionization potential (FIP) order of the elements to examine whether the well
known FIP or inverse FIP effect is present in Capella. Based on the RGS fit, one
may tentatively conclude that the Capella abundances appear to manifest the
solar-like FIP effect. However, we note that our derived abundances are
significantly different from earlier measurement of \citet{audard03} using
the same RGS data and a four-temperature fitting model. Moreover, there are a
few problems in the DEM and abundance determinations 
using the RGS data alone, especially due to the calibration uncertainties in
the short wavelength region, which may bias the results. In our analysis of
EPIC and RGS spectra of some strong sources, we discovered that the RGS
effective area below 8~{\AA} is underestimated by $\sim 20$\% relative to
EPIC. To make up this loss of flux, the DEM derived from RGS data show a
significant high temperature tail, and the derived Si abundance may have also
been overestimated. 

\section{\chandra HETGS}
\label{sec:hetg}
We use the same model as in the RGS fit for the analysis of the \chandra
HETGS data. The $\pm 1$ order spectra of both MEG and HEG are jointly
fit. Because HETGS has very little effective area at the wavelengths of C
K-shell lines, the C abundance is fixed at the value derived from the RGS fit. The
reconstructed DEM and abundances are shown in Figures~\ref{fig:dem} and
\ref{fig:ab} as the red histogram and red symbols, respectively. Both orders of
MEG spectra are summed and 
shown in Figure~\ref{fig:hetgsp} for the spectral region between 6 and
18~{\AA}. The model and data spectra are seen to agree with each other very
well. As compared with the RGS determination, the total emission
measure in the two peaks is larger and the abundances smaller on average. This
is clearly due to the 
same reason we discussed in our two test cases, i.e., the weak continuum emission
does not constrain the total emission
measure and absolute abundance values well, although relative abundances are
reasonably well determined. The discrepancy in the total emission
measure derived from the two datasets are much larger than that seen in the
test cases. It is possible that the uncertainties in the atomic database,
instrument calibration, and background many have caused larger differences.
Another major difference in the DEM is that the
secondary peak appears to be much stronger relative to 
the main peak than in the RGS fit, and slightly shifted to higher
temperatures. However, this peak is poorly 
constrained in the HETGS data because it does not have significant effective
area for the intermediate-Z L-shell lines, and we believe that the RGS derived
properties for this temperature region are more robust. For temperatures above
$\log(T)=7.4$, the HETGS fit shows no emission measure while RGS fit gives some
significant values. Because HETGS has better sensitivity at shorter wavelengths,
and we have reason to believe that the RGS effective area below 8~{\AA} is
underestimated, we conclude that the HETGS fit is more reliable than the RGS
fit in this temperature region.

\section{Combined Analysis of RGS and HETGS}
\label{sec:rhetg}
The previous sections show that the RGS data is better in constraining the low
temperature part of the DEM while the HETGS data is
more suitable to constrain the high temperature part of the DEM. To combine
the strengths of both instruments, we have carried out a joint fit of both
datasets using the same model. Even though the two observations
are not made simultaneously, the lack of variability of the Capella corona
makes such an effort possible. It is also justified by the fact that the joint
fit model describes both HETGS and RGS spectra very well. More quantitively
speaking, the likelyhood statistics of the goodness of fit, $X^2$, for the
model derived 
from the HETGS data is 29891; that for the model derived from the RGS
data is 38489; when the model derived from the joint fit is applied to
MEG and RGS data, the $X^2$ values are 30085 and 38527, respectively. They are all
slightly larger than the values for the individual fits, but within the
expected statistical deviations. To avoid the RGS
calibration problems below 8~{\AA}, we ignore that region in the fit since
HETGS already provides adequate statistics at shorter wavelengths. The resulting DEM
reconstruction and the abundance measurements are shown in
Figures~\ref{fig:dem} and \ref{fig:ab} as the black histogram with error bars
and the black symbols, respectively. As expected, the secondary peak
in the joint fit model closely resembles that in the RGS fit since it is mainly
constrained by the intermediate-Z L-shell lines, which are only present in the
RGS data, and the derived N abundance is also closer to that obtained from
the RGS fit. The high temperature tail in the DEM is significantly suppressed as
compared with the RGS fit due to the constraints provided by the HETGS data. To
make up the lost continuum in the RGS wavelength band due to this high temperature
tail, the main peak in the DEM is made slightly larger and broader. As a
result, the abundances determined from the joint fit are slightly smaller
than the RGS result. It is also seen that the uncertainties are smaller
than either the RGS or the HETGS derived ones. We believe that the result of
this joint fit represents the best model for the X-ray emission of the Capella
corona. We note that derived abundances in our joint fit model still suggest a
solar-like FIP 
effect. Overall, the abundances of elements with FIP less than 10 eV are
slightly above solar, while those of elements with FIP greater than 10 eV are
below the solar values. The most notable exception to this trend is the N
abundance, which is above solar despite of having a 14.5 eV FIP. 

\begin{figure}
\epsscale{1.0}
\plotone{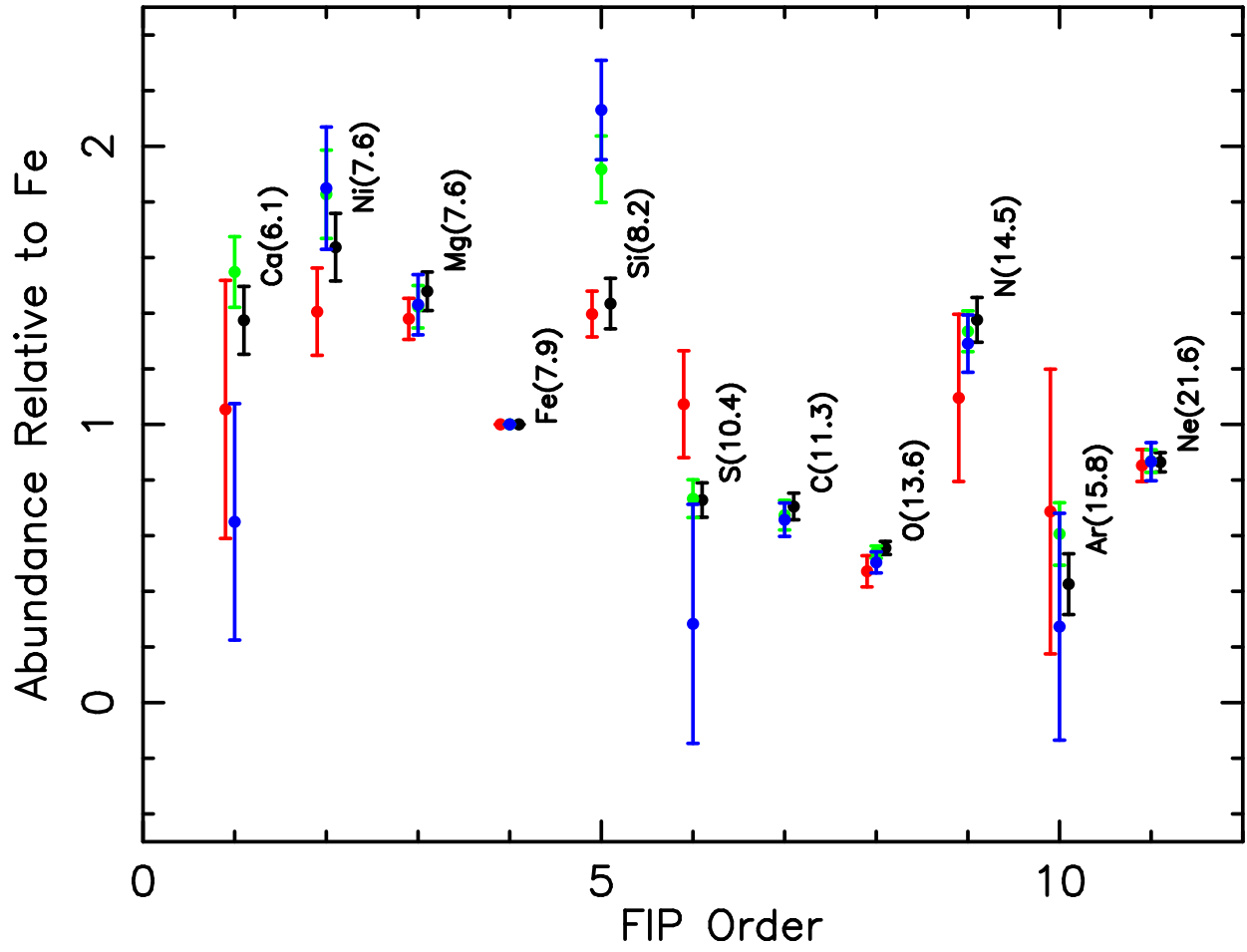}
\caption{\label{fig:abr} Same as in Figure~\ref{fig:ab}, except that
  abundances are relative to that of Fe.}
\end{figure}

One might be a little concerned that the abundances derived from individual
analyses of RGS and HETGS data do not agree with each other within the stated
uncerainties. However, as we have 
already pointed out earlier, this is mainly the result of a weak continuum,
which makes the absolute determination of abundances difficult. The
uncertainties in our measurements are obtained by taking the variance along the
MCMC chain in equilibrium. However, as we discussed earlier, we do not iterate
the overall normalization parameter, $\eta$, in our analysis. Therefore it is
not surprising to see very different absolute abundance values from different
analyses. The relative abundances are still well constrained in our method. In
Figure~\ref{fig:abr}, we plot 
the same data of Figure~\ref{fig:ab} with abundances measured relative to
that of Fe. It is seen that measurements from different data sets are generally
consistent with each other. Si is the only exception, where the RGS derived value
is significantly larger than that from HETGS. We atrribute this difference to
the RGS calibration problems near the Si K-shell lines.

To facilitate easy comparison with other analyses, we tabulate the DEMs and
abundances in our various reconstructions in Table~\ref{tab:dem} and
Table~\ref{tab:ab}. 

\section{\chandra LETGS}
\label{sec:letg}
Since the \chandra LETGS data has even longer wavelength coverage than the
RGS, we would expect that it provides even more stringent constraints on
the low temperature DEM. However, the atomic data beyond about 40~{\AA} is
not as well known as those at shorter wavelengths, and it is possible that
the LETGS effective area calibration may be problematic at longer wavelengths.
This may be partially blamed for the large discrepancies found between the
theoretical and observed ratios of EUV to X-ray line intensities of Fe XVIII
and XIX discovered by \citet{desai05}, 
which we will discuss later in more detail. Another complication of fitting
the LETGS raw spectra is that its wavelengths scale is not as well determined
as for RGS and HETGS because of the non-linearity of the HRC-S
detector. Therefore, a better method for analyzing the LETGS data is to first
identify and measure strong lines, which can then be used for DEM
analysis. \citet{argiroffi03} adopted such an approach and used an MCMC DEM
deconvolution method which is different from the one we developed here. Their
derived DEM and abundances are quite similar to our results derived from the
joint fit of HETGS and RGS data. However, because only a limited set of
unblended lines can be used in their method, the associated uncertainties on
abundances are much larger than our determination. 

\begin{figure}
\epsscale{1.0}
\plotone{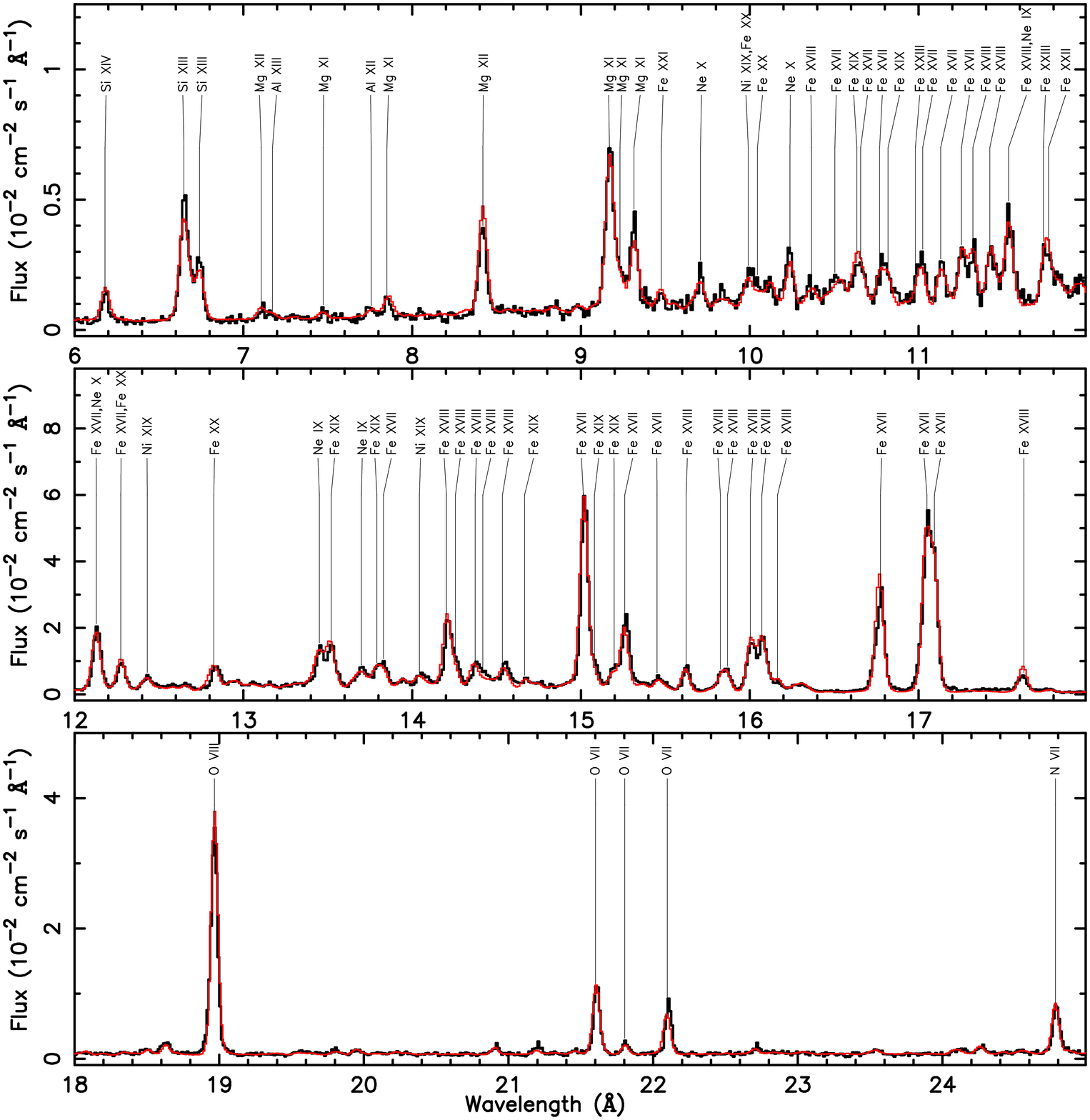}
\caption{\label{fig:letgsp}The comparison of measured and modeled LETGS
  spectrum in the 6--25~{\AA} region. The black line is the sum of $\pm 1$
  orders of LETGS spectra; the red line is the model using the DEM and
  abundances reconstructed from the joint fit of HETGS and RGS data.}
\end{figure}

Due to the lack of variability of the Capella emission, we expect the model
derived from the RGS and HETGS data to fit the LETGS X-ray spectra equally well. In
the present analysis, we simply verify that this is the case without carrying
out a full DEM analysis on the LETGS data. By folding the RGS and HETGS joint
fit model through the LETGS response, we show the comparison of the calculated and
observed spectra in Figure~\ref{fig:letgsp} for the wavelength region between 6
and 25~{\AA}. It is clear that the model is a very good representation of all
significant line and continuum emission. 

Such good agreement is lost for wavelengths above 40~{\AA}, as we have noted
that the theoretical atomic data may be more uncertain, and there may be
emission regions with lower temperatures 
which are not constrained by either RGS or HETGS. In the present work, we do
not investigate the many 
discrepancies found in the long wavelength range. However, we further
study the problem identified by \citet{desai05}, i.e., the observed intensity
ratios of EUV to X-ray lines of Fe XVIII and XIX are found to be a factor of 2
smaller than calculations. Here we expand the analysis to include more
Fe L-shell ions.

\begin{figure}
\epsscale{1.0}
\plotone{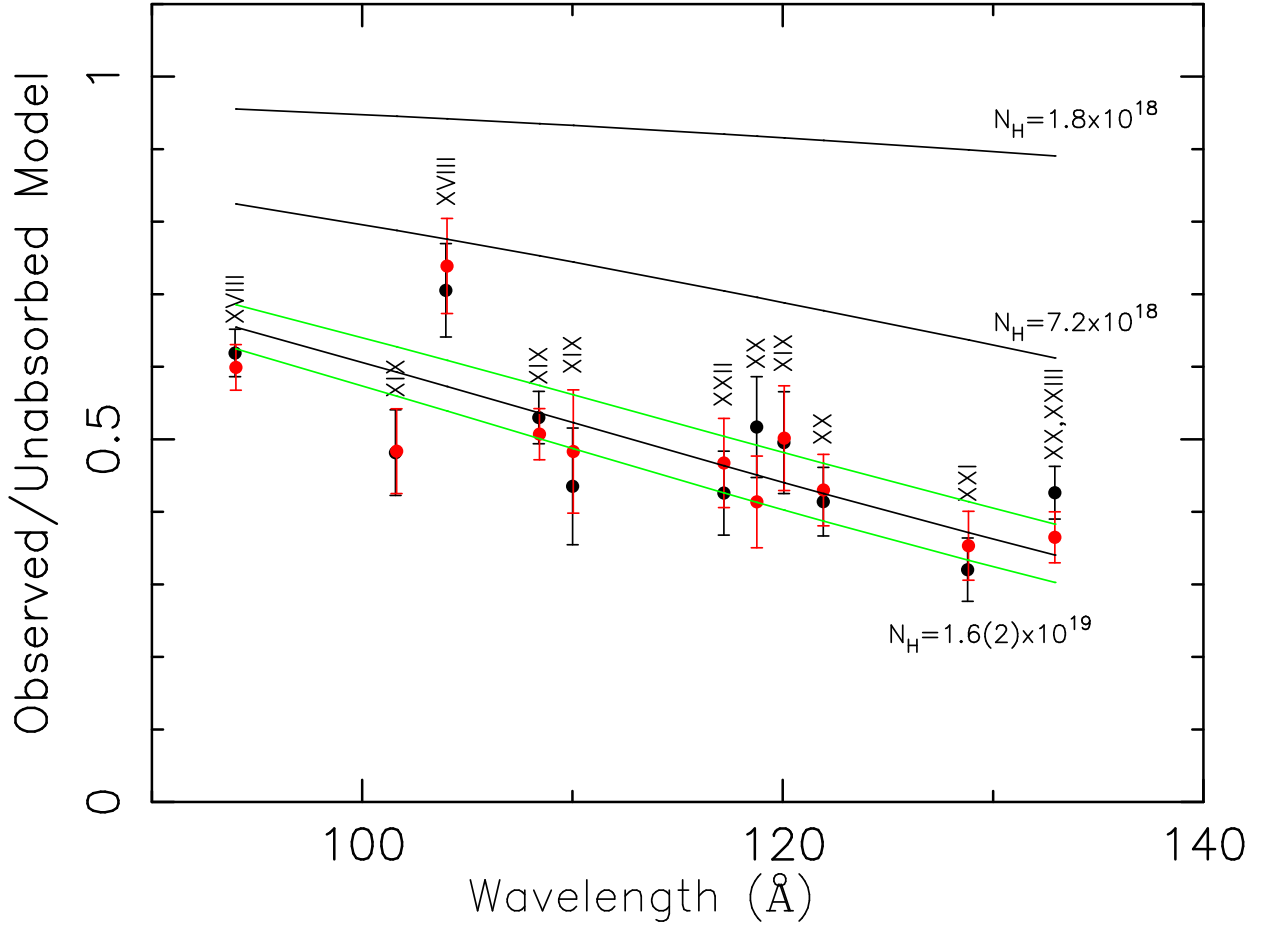}
\caption{\label{fig:letgflux}The comparison of observed and modeled fluxes of
  EUV lines of Fe XVIII--XXIII in the 90--140~{\AA} region. The black symbols
  are the ratio of measured line fluxes from $+1$ order LETGS spectrum to the
  unabsorbed model, where the model is calculated with the DEM and abundances
  reconstructed from the joint fit of HETGS and RGS data; the red symbols
  are measurements from the $-1$ order spectrum. The bottom black line and the
  two green lines that fit the
  data points indicate the effective neutral hydrogen column density, $1.6\pm
  0.2\times 10^{19}$ ~cm$^{-2}$, required to bring the model in agreement with
  the data. The top black line indicates the absorption provided by the true
  neutral hydrogen column density of $1.8\times 10^{18}$~cm$^{-2}$
  \citep{linsky93}. The middle black line indicates the absorption provided by
  the effective neutral hydrogen column density of $7.2\times
  10^{18}$~cm$^{-2}$, assuming H II ionization fraction of 0.8 and He II
  fraction of 0.5 for a partially ionized interstellar medium.}
\end{figure}

In Table~\ref{tab:letgflux}, we list the measured and modeled fluxes of
several lines in the 90 -- 140~{\AA} region, where most EUV lines of
Fe ions are located. 
These lines all originate from the $n=2\to 2$ transitions of L-shell ions. It
is clear that all lines are overestimated in the model by at least a factor of
two. Extracting the individual line intensities in the data and the unabsorbed
model, we plot the ratio as a function of wavelength in
Figure~\ref{fig:letgflux}. A total of 11 line 
features from Fe XVIII--XXIII are used, some of them are blends of more than
one line, in which case the model intensity includes all contributions. The
correction factor due to the interstellar absorption corresponding to the
neutral hydrogen column density of $N_H = 1.8\times 10^{18}$~cm$^{-2}$ is also
shown, which is a very small effect. The data points suggest a trend as if the
absorption column density is much higher, assuming that the LETGS calibration at
these wavelengths relative to the X-ray band is correct. The
effective $N_H$ must be increased to $1.6\pm0.2 \times 10^{19}$~cm$^{-2}$ in
order to explain the observed data. 

The neutral hydrogen column density of $N_H=1.8\times 10^{18}$~cm$^{-2}$ is
measured using the interstellar absorption profile of the hydrogen Ly$\alpha$
line toward Capella, and has a very small uncertainty
\citep{linsky93}. However, the hydrogen 
ionization fraction of the local intersetallar cloud is quite uncertain. Because 
the absorption model used in our analysis and that of \citet{desai05} assumes
that the interstellar medium is cold and neutral with standard cosmic
abundances for all elements, the true neutral
hydrogen column density may not represent all of the absorbing
materials. At the wavelength region of interest, the most important
contributors of absorption are H I, He I, and He II. 
There are considerable evidences that hydrogen and helium in the local
interstellar cloud are partially ionized with a H II fraction of 0.3--0.8
\citep{wood02}. The detailed photoionization model of \citet{slavin02} gives
the H II fraction of 0.2--0.3, and He II fraction of 0.3--0.5 with little He
III. Let us define an effective neutral hydrogen column density, $\tilde{N}_H$
for a partially ionized medium, such that the absorption at certain photon
energy in this medium is the same as a neutral one with column density
equal to $\tilde{N}_H$, i.e.,
\begin{equation}
\tilde{N}_H\left(\sigma(H I) + A_{He}\sigma(He I)\right) = N(H I)\sigma(H I) +
N(He I)\sigma(He I) + N(He II)\sigma(He II),
\end{equation}
where $A_{He} = 0.1$ is the cosmic He abundance. Assuming that the H II fraction
is X(H), He II fraction is X(He), and no He III,
\begin{equation}
\tilde{N}_H = N(H I)\frac{[1-X(H)]\sigma(H I) + A_{He}[1-X(He)]\sigma(He I) +
  A_{He}X(He)\sigma(He II)}{[1-X(H)][\sigma(H I) + A_{He}\sigma(He I)]}.
\end{equation}
For photoabsoption cross sections near 100~{\AA}, $\sigma(He I) \sim 22\sigma(H
I)$, and $\sigma(He II) \sim 16\sigma(H I)$, and therefore
\begin{equation}
\tilde{N}_H = N(H I) \frac{3.3 - X(H) + 0.6X(He)}{3.3[1-X(H)]},
\end{equation}
where we have used $A_{He} = 0.1$. Clearly, it depends strongly on $X(H)$ but
weakly on $X(He)$. Assuming the average value of $X(He)=0.5$ and $N(H I) =
1.8\times 10^{18}$~cm$^{-2}$, one requires $X(H) = 0.91$ in order to obtain
$\tilde{N}_H = 1.6\times 10^{19}$~cm$^{-2}$. This H II fraction is outside the
range estimated by \citet{wood02}, which is based on the measurement of number
densities of neutral hydrogen and electrons. Taking the most extreme value of
$X(H)=0.8$ from the observation would imply $\tilde{N}_H=7.2\times
10^{18}$~cm$^{-2}$, which is only 
enough to account for about half of the discrepancies. To explain the
remaining disagreement, one must assume that either the theoretical ratios of
EUV to X-ray lines of Fe L-shell ions are too large by about 50\%,  the LETG
effective area calibration is off by the same amount, or the combination of
both. Either of these cases seems plausible, especially the last one, since
one would only need to assume systematic errors of 20--30\% in both factors.

\section{Discussion and Conclusions}
\label{sec:discussion}
During this reanalysis of Capella data, we developed a new atomic line
list based on APED/APEC. The improvements include better ionization
balance calculations for all L-shell ions, more level population mechanisms
for Fe L-shell ions such as resonance excitation, recombination, and
ionization processes, better wavelengths based on either accurate many-body
perturbation theory or laboratory measurements for L-shell lines of S,
Ar, Fe, and Ni. We have tested the new line database on many other coronal
sources, and found it to always perform better than APED/APEC. However, there
are still some serious deficiencies in the atomic data. For example, the
L-shell lines of elements other than Fe only include direct excitation as the
level population mechanism; the line emissivities are calculated in the low
density limit; and the wavelengths of high-$n$ transitions of Fe L-shell ions
have not been well determined as those of $3\to 2$ transitions. Some of these
problems are being addressed with the ongoing laboratory and theoretical
work. We expect to further improve the database in the near future.

We also developed a new MCMC based DEM deconvolution method. It is different
from existing methods in several aspects, including the parameterization of
DEM and the utilization of the raw spectra instead of line fluxes. We
demonstrated that the HETGS data alone often cannot constrain the total
emission measure and the absolute abundances independently due to the lack of
significant continuum emission. The DEM derived from RGS alone often has a
high temperature tail which may bias the abundance measurements. The
combination of HETGS and RGS data were shown to give the most robust results. 
The abundances derived from the joint analysis of HETGS and RGS indicate the
presence of solar-like FIP effect. 

We investigated the problem of intensity ratios of $2\to 2$ EUV lines to
$3\to 2$ X-ray lines of Fe L-shell ions using the LETGS data. The
overestimation of $2\to 2$ lines are shown to exist not only for Fe XVIII and
XIX, but also for higher charge ions up to Fe XXIII. The discrepancy appears
to grow larger for lines at longer wavelengths. We proposed a partial
explanation by assuming that the interstellar medium along the Capella line
sight is partially ionized, and therefore has a larger effective absorption
column density than if the medium is neutral. However, constraints on the
ionization fraction of H II dictate that it can only account for about half
of the discrepancy. It is plausible that the remaining discrepancy is due to
either the LETGS calibration uncertainties or systematic errors in theoretical
line ratios.

\acknowledgments
The authors acknowledge the support by the NASA grants NAG5-5419 and
NNGG04GL76G. 


\begin{thebibliography}{29}
\expandafter\ifx\csname natexlab\endcsname\relax\def\natexlab#1{#1}\fi

\bibitem[{Anders \& Grevesse}(1989)Anders \& Grevesse]{anders89}
Anders, E. \&  Grevesse, N. 1989, Geochim. Cosmochim. Acta, 53, 197

\bibitem[{Argiroffi {et~al.}(2003)Argiroffi, Maggio, \& Peres}]{argiroffi03}
Argiroffi, C., Maggio, A., \& Peres, G. 2003, \aap, 404, 1033

\bibitem[{Audard {et~al.}(2001)Audard {et~al.}}]{audard01}
Audard, M. et~al. 2001, \aap, 365, L329

\bibitem[{Audard {et~al.}(2003)Audard, G\"{u}del, Sres, Raassen, \&
  Mewe}]{audard03}
Audard, M., G\"{u}del, M., Sres, A., Raassen, A. J.~J., \& Mewe, R. 2003, \aap,
  398, 1137

\bibitem[{Behar {et~al.}(2001)Behar, Cottam, \& Kahn}]{behar01}
Behar, E., Cottam, J., \& Kahn, S.~M. 2001, \apj, 548, 966

\bibitem[{Brickhouse {et~al.}(2000)Brickhouse, Dupree, Edgar, Liedahl, Drake,
  White, \& Singh}]{brickhouse00}
Brickhouse, N.~S., Dupree, A.~K., Edgar, R.~J., Liedahl, D.~A., Drake, S.~A.,
  White, N.~E., \& Singh, K.~P. 2000, \apj, 530, 387

\bibitem[{Brinkman {et~al.}(2000)Brinkman {et~al.}}]{brinkman00}
Brinkman, A.~C. et~al. 2000, \apj, 530, L111

\bibitem[{Brown {et~al.}(1998)Brown, Beiersdorfer, Liedahl, Widmann, \&
  Kahn}]{brown98}
Brown, G.~V., Beiersdorfer, P., Liedahl, D.~A., Widmann, K., \& Kahn, S.~M.
  1998, \apj, 502, 1015

\bibitem[{Brown {et~al.}(2002)Brown, Beiersdorfer, Liedahl, Widmann, Kahn, \&
  Clothiaux}]{brown02}
Brown, G.~V., Beiersdorfer, P., Liedahl, D.~A., Widmann, K., Kahn, S.~M., \&
  Clothiaux, E.~J. 2002, \apjs, 140, 589

\bibitem[{Canizares {et~al.}(2000)Canizares {et~al.}}]{canizares00}
Canizares, C.~A. et~al. 2000, \apj, 539, L41

\bibitem[{Desai {et~al.}(2005)Desai {et~al.}}]{desai05}
Desai, P. et~al. 2005, \apj, 625, L59

\bibitem[{Doron \& Behar(2002)}]{doron02}
Doron, R. \& Behar, E. 2002, \apj, 574, 518

\bibitem[{Dupree {et~al.}(1993)Dupree, Brickhouse, Doschek, Green, \&
  Raymond}]{dupree93}
Dupree, A.~K., Brickhouse, N.~S., Doschek, G.~A., Green, J.~C., \& Raymond,
  J.~C. 1993, \apj, 418, L41

\bibitem[{Gu(2003a)}]{gu03a}
Gu, M.~F. 2003a, \apj, 582, 1241

\bibitem[{Gu(2003b)}]{gu03b}
---. 2003b, \apj, 590, 1131

\bibitem[{Gu(2005)}]{gu05}
---. 2005, \apjs, 156, 105

\bibitem[{Hummel {et~al.}(1994)Hummel, Armstrong, Quirrenbach, Buscher,
  Mozurkewich, \& Elias}]{hummel94}
Hummel, C.~A., Armstrong, J.~T., Quirrenbach, A., Buscher, D.~F., Mozurkewich,
  D., \& Elias, N.~M. 1994, \aj, 107, 1859

\bibitem[{Kashyap \& Drake(1998)}]{kashyap98}
Kashyap, V. \& Drake, J.~J. 1998, \apj, 503, 450

\bibitem[{Lepson {et~al.}(2003)Lepson, Beiersdorfer, Behar, \& Kahn}]{lepson03}
Lepson, J.~K., Beiersdorfer, P., Behar, E., \& Kahn, S.~M. 2003, \apj, 590, 604

\bibitem[{Lepson {et~al.}(2005)Lepson, Beiersdorfer, Behar, \& Kahn}]{lepson05}
---. 2005, \apj, 625, 1045

\bibitem[{Linsky {et~al.}(1993)Linsky {et~al.}}]{linsky93}
Linsky, J.~L. et~al. 1993, \apj, 402, 694

\bibitem[{Manset {et~al.}(2000)Manset, Veillet, \& Crabtree}]{houck00}
Manset, N., Veillet, C., \& Crabtree, D., eds. 2000, 591

\bibitem[{Mazzotta {et~al.}(1998)Mazzotta, Mazzitelli, Colafrancesco, \&
  Vittorio}]{mazzotta98}
Mazzotta, P., Mazzitelli, G., Colafrancesco, S., \& Vittorio, N. 1998, \aaps,
  133, 403

\bibitem[{Mewe {et~al.}(2001)Mewe, Raassen, Drake, Kaastra, {van der Meer}, \&
  Porwuet}]{mewe01}
Mewe, R., Raassen, A. J.~J., Drake, J.~J., Kaastra, J.~S., {van der Meer}, R.
  L.~J., \& Porwuet, D. 2001, \aap, 368, 888

\bibitem[{Peterson {et~al.}(2006)Peterson, Marshall, \& Andersson}]{peterson06}
Peterson, J.~R., Marshall, P.~J., \& Andersson, K. 2006, \apj, submitted

\bibitem[{Phillips {et~al.}(2001)Phillips, Mathioudakis, Huenemoerder,
  Williams, Phillips, \& Keenan}]{phillips01}
Phillips, K. J.~H., Mathioudakis, M., Huenemoerder, D.~P., Williams, D.~R.,
  Phillips, M.~E., \& Keenan, F.~P. 2001, \mnras, 325, 1500

\bibitem[{Slavin \& Frisch(2002)}]{slavin02}
Slavin, J.~D. \& Frisch, P.~C. 2002, \apj, 565, 364

\bibitem[{Smith {et~al.}(2001)Smith, Brickhouse, Liedahl, \& Raymond}]{smith01}
Smith, R.~K., Brickhouse, N.~S., Liedahl, D.~A., \& Raymond, J.~C. 2001, \apj,
  556, 91

\bibitem[{Wood {et~al.}(2002)Wood, Redfield, Linsky, \& Sahu}]{wood02}
Wood, B.~E., Redfield, S., Linsky, J.~L., \& Sahu, M.~S. 2002, \apj, 581, 1168

\end{thebibliography}

\clearpage
\begin{deluxetable}{ ccccc }
\tabletypesize{\scriptsize}
\tablecaption{\label{tab:dem}Reconstructed DEMs of the Capella corona in units
  of $4\pi D^2\times 10^{12}$ cm$^{-3}$, where $D$ is the distance of the
  source. The numbers in parentheses are 1$\sigma$ statistical uncertainties.}
\tablehead{
\colhead{$\log(T/\mbox{K})$} &
\colhead{RGS\tablenotemark{a}}&
\colhead{RGS-L\tablenotemark{b}} &
\colhead{HETGS\tablenotemark{c}} &
\colhead{RGS+HETGS\tablenotemark{d}}
}
\startdata
5.825 &  2.43(1.11) &  2.55(1.44) &  8.67(3.86) &  1.24(1.04) \\
5.875 &  1.63(1.01) &  1.98(1.30) &  7.94(3.60) &  1.00(0.91) \\
5.925 &  0.90(0.78) &  1.20(0.93) &  7.40(3.62) &  0.57(0.67) \\
5.975 &  0.46(0.57) &  0.60(0.66) &  6.47(3.51) &  0.36(0.59) \\
6.025 &  0.31(0.48) &  0.26(0.44) &  5.12(3.18) &  0.29(0.51) \\
6.075 &  0.26(0.45) &  0.14(0.32) &  3.97(2.73) &  0.22(0.42) \\
6.125 &  0.23(0.42) &  0.07(0.24) &  2.65(2.29) &  0.27(0.47) \\
6.175 &  0.33(0.50) &  0.09(0.26) &  1.67(1.90) &  0.41(0.59) \\
6.225 &  0.56(0.57) &  0.19(0.38) &  1.32(1.65) &  0.59(0.67) \\
6.275 &  0.87(0.76) &  0.65(0.63) &  1.27(1.71) &  1.20(0.93) \\
6.325 &  1.16(0.81) &  1.30(0.85) &  1.74(1.88) &  1.92(1.08) \\
6.375 &  1.45(0.98) &  2.33(1.12) &  3.02(2.45) &  1.68(1.15) \\
6.425 &  1.66(1.05) &  3.10(1.44) &  6.22(3.65) &  1.54(1.02) \\
6.475 &  1.27(0.95) &  2.48(1.40) &  8.74(3.63) &  1.46(1.05) \\
6.525 &  1.41(1.03) &  2.01(1.19) & 11.32(4.38) &  1.71(1.09) \\
6.575 &  2.31(1.33) &  1.93(1.17) & 10.74(4.38) &  2.91(1.47) \\
6.625 &  3.55(1.42) &  2.38(1.27) & 10.71(4.31) &  4.40(1.75) \\
6.675 &  5.10(1.73) &  3.90(1.65) & 10.39(4.39) &  6.76(2.13) \\
6.725 &  6.02(1.87) &  5.81(1.87) & 11.43(4.64) &  8.17(2.37) \\
6.775 &  6.89(2.09) &  7.75(2.12) & 13.49(4.64) &  8.37(2.42) \\
6.825 &  7.18(2.06) &  7.92(2.14) & 15.76(5.08) &  8.65(2.56) \\
6.875 &  6.99(1.88) &  7.24(2.16) & 19.25(5.36) &  9.55(2.38) \\
6.925 &  6.83(1.83) &  6.21(1.97) & 18.90(4.97) &  9.27(2.43) \\
6.975 &  6.17(1.73) &  5.89(1.80) & 14.00(4.36) &  7.77(2.11) \\
7.025 &  3.48(1.32) &  3.37(1.37) &  9.12(3.60) &  4.57(1.74) \\
7.075 &  1.18(0.85) &  1.46(0.96) &  5.76(2.83) &  2.02(1.25) \\
7.125 &  0.55(0.58) &  0.72(0.67) &  3.01(2.29) &  1.34(0.93) \\
7.175 &  0.32(0.50) &  0.41(0.55) &  1.49(1.49) &  1.14(0.89) \\
7.225 &  0.34(0.50) &  0.23(0.41) &  0.51(0.95) &  0.79(0.77) \\
7.275 &  0.42(0.55) &  0.29(0.47) &  0.17(0.59) &  0.72(0.74) \\
7.325 &  0.36(0.52) &  0.36(0.52) &  0.01(0.18) &  0.56(0.67) \\
7.375 &  0.37(0.55) &  0.39(0.54) &  0.00(0.00) &  0.34(0.54) \\
7.425 &  0.40(0.56) &  0.35(0.54) &  0.00(0.00) &  0.26(0.50) \\
7.475 &  0.41(0.56) &  0.43(0.59) &  0.00(0.00) &  0.27(0.49) \\
7.525 &  0.53(0.62) &  0.58(0.68) &  0.00(0.00) &  0.32(0.54) \\
7.575 &  0.77(0.78) &  0.89(0.80) &  0.00(0.00) &  0.36(0.57) \\
7.625 &  1.22(0.96) &  1.17(1.00) &  0.00(0.00) &  0.52(0.67) \\
7.675 &  1.82(1.14) &  1.69(1.15) &  0.00(0.00) &  0.55(0.66) \\
7.725 &  2.46(1.26) &  2.46(1.39) &  0.00(0.00) &  0.74(0.78) \\
7.775 &  3.90(1.48) &  3.69(1.48) &  0.00(0.00) &  0.99(0.81) \\
\enddata

\tablenotetext{a}{Reconstruction from the RGS data alone.}
\tablenotetext{b}{Reconstruction from the RGS data alone and exlucding
  intermediate-Z L-shell lines.}
\tablenotetext{c}{Reconstruction from the HETGS data alone.}
\tablenotetext{d}{Reconstruction from the joint RGS and HETGS data.}

\end{deluxetable}

\clearpage
\begin{deluxetable}{ cccccc }
\tablecaption{\label{tab:ab}Measured abundances in solar photospheric units of
  \citet{anders89}. The numbers in parentheses are statistical
  uncertainties at 90\% confidence level.}
\tablehead{
\colhead{$Z$} &
\colhead{FIP\tablenotemark{a} (eV)} &
\colhead{RGS\tablenotemark{b}}&
\colhead{RGS-L\tablenotemark{c}} &
\colhead{HETGS\tablenotemark{d}} &
\colhead{RGS+HETGS\tablenotemark{e}}
}
\startdata
 6 & 11.3 & 0.69(0.05) & 0.68(0.06) & $\cdots$ & 0.57(0.04) \\
 7 & 14.5 & 1.38(0.08) & 1.33(0.11) & 0.49(0.05) & 1.10(0.06) \\
 8 & 13.6 & 0.56(0.02) & 0.52(0.04) & 0.21(0.08) & 0.45(0.02) \\
10 & 21.6 & 0.89(0.04) & 0.90(0.07) & 0.38(0.02) & 0.69(0.03) \\
12 &  7.6 & 1.47(0.08) & 1.48(0.11) & 0.62(0.04) & 1.19(0.06) \\
14 &  8.2 & 1.98(0.12) & 2.20(0.18) & 0.62(0.08) & 1.15(0.07) \\
16 & 10.4 & 0.76(0.07) & 0.29(0.44) & 0.48(0.12) & 0.58(0.05) \\
18 & 15.8 & 0.63(0.12) & 0.28(0.42) & 0.31(0.07) & 0.34(0.09) \\
20 &  6.1 & 1.60(0.13) & 0.67(0.44) & 0.47(0.12) & 1.10(0.10) \\
26 &  7.9 & 1.03(0.03) & 1.03(0.08) & 0.45(0.13) & 0.80(0.03) \\
28 &  7.6 & 1.88(0.16) & 1.91(0.23) & 0.63(0.03) & 1.31(0.10) \\
\enddata

\tablenotetext{a}{First ionization potential.}
\tablenotetext{b}{Reconstruction from the RGS data alone.}
\tablenotetext{c}{Reconstruction from the RGS data alone and exlucding
  intermediate-Z L-shell lines.}
\tablenotetext{d}{Reconstruction from the HETGS data alone.}
\tablenotetext{e}{Reconstruction from the joint RGS and HETGS data.}

\end{deluxetable}

\clearpage
\begin{deluxetable}{ ccccc }
\tablecaption{\label{tab:letgflux}Observed and modeled fluxes of EUV lines of
  Fe XVIII--XXIII in the 90--140~{\AA} region. The observed values are from
  $\pm$ orders of LEG spectra. The numbers in parentheses are 1$\sigma$
  statistical uncertainties.}
\tablehead{
\colhead{Ion} &
\colhead{$\lambda$ ({\AA})} &
\multicolumn{3}{c}{Flux (photons cm$^{-2}$ ks$^{-1}$)} \\
\colhead{} &
\colhead{} &
\colhead{$-1$ order} &
\colhead{$+1$ order} &
\colhead{Model\tablenotemark{a}}
}
\startdata
Fe XVIII &  94.0 & 3.96(0.10) & 3.83(0.10) & 6.40 \\
Fe XIX & 101.6 & 0.90(0.05) & 0.90(0.05) & 1.86 \\
Fe XVIII & 104.0 & 1.65(0.08) & 1.73(0.08) & 2.33 \\
Fe XIX & 108.4 & 2.98(0.10) & 2.86(0.10) & 5.63 \\
Fe XIX & 110.0 & 0.41(0.04) & 0.46(0.04) & 0.95 \\
Fe XXII & 117.2 & 0.78(0.05) & 0.85(0.06) & 1.82 \\
Fe XX & 118.8 & 0.80(0.05) & 0.64(0.05) & 1.56 \\
Fe XIX & 120.1 & 0.75(0.05) & 0.76(0.05) & 1.51 \\
Fe XX & 121.9 & 1.26(0.07) & 1.31(0.08) & 3.04 \\
Fe XXI & 128.8 & 1.28(0.09) & 1.41(0.10) & 4.00 \\
Fe XX,XXIII & 132.9 & 3.22(0.14) & 2.76(0.13) & 7.56 \\
\enddata

\tablenotetext{a}{The model fluxes are derived from
  the DEM and abundances reconstructed from the joint fit of HETGS and RGS
  data without including corrections due to interstellar absorption.}

\end{deluxetable}

\clearpage

\end{document}